\documentclass[preprint,12pt]{elsarticle}

\usepackage{amssymb}
\usepackage{setspace}        
\usepackage{multirow}
\usepackage{bm}
\usepackage{color}

\usepackage{booktabs}
\usepackage{subfig}
\usepackage{graphicx}
\usepackage{amsmath}

\journal{Physica A: Statistical Mechanics and its Applications}

\begin{document}
	
\begin{spacing}{1.5}

\begin{frontmatter}
	


\title{Dynamic Correlation of Market Connectivity, Risk Spillover and Abnormal Volatility in Stock Price}

\author[a]{Muzi Chen}
\author[b]{Nan Li}
\author[a]{Lifen Zheng}
\author[c,*]{Difang Huang}
\author[d,*]{Boyao Wu}

\affiliation[a]{
	organization={School of Management Science and Engineering, Central University of Finance and Economics},
	postcode={102206}, 
	state={Beijing},
	country={China}		
}
\affiliation[b]{
	organization={Business School, Shandong Normal University},
	postcode={250014}, 
	state={Shandong},
	country={China}		
}			
\affiliation[c]{
	organization={Department of Econometrics and Business Statistics, Monash University},
	city={Melbourne},
	postcode={3145}, 
	state={Victoria},
	country={Australia}		
}
\affiliation[d]{
	organization={School of Banking and Finance, University of International Business and Economics},
	postcode={100029}, 
	state={Beijing},
	country={China}		
}

\affiliation[*]{Corresponding author: Difang Huang, difang.huang@monash.edu; Boyao Wu, boyao581@gmail.com}

\begin{abstract}
	The connectivity of stock markets reflects the information efficiency of capital markets and contributes to interior risk contagion and spillover effects. We compare Shanghai Stock Exchange A-shares (SSE A-shares) during tranquil periods, with high leverage periods associated with the 2015 subprime mortgage crisis. We use Pearson correlations of returns, the maximum strongly connected subgraph, and $3\sigma$ principle to iteratively determine the threshold value for building a dynamic correlation network of SSE A-shares. Analyses are carried out based on the networking structure, intra-sector connectivity, and node status, identifying several contributions. First, compared with tranquil periods, the SSE A-shares network experiences a more significant small-world and connective effect during the subprime mortgage crisis and the high leverage period in 2015. Second, the finance, energy and utilities sectors have a stronger intra-industry connectivity than other sectors. Third, HUB nodes drive the growth of the SSE A-shares market during bull periods, while stocks have a think-tail degree distribution in bear periods and show distinct characteristics in terms of market value and finance. Granger linear and non-linear causality networks are also considered for the comprison purpose. Studies on the evolution of inter-cycle connectivity in the SSE A-share market may help investors improve portfolios and develop more robust risk management policies.

\end{abstract}


\begin{highlights}
	\item A novel threshold determination method is proposed for constructing dynamic networks.
	\item The dynamic evolution of the Shanghai Stock Exchange A-shares market is analyzed from multiple views.
	\item Factors driving the connectivity evolution of industries and stocks are investigaed.
	\item We develop the dynamic model of HUB nodes and discover the significant impact of market values on stock statuses.
\end{highlights}


\end{frontmatter}



\section{Introduction}
\label{}


The Shanghai Stock Exchange A-shares (SSE A-shares) capital market has made considerable progress over the past three decades. On the one hand, stocks in the SSE A-shares market show increasing connectivity due to improved information transmission efficiency, a rise in cross-shareholding, and deeper relationships with overseas markets. On the other hand, the SSE A-shares market, as an emerging market, is highly volatile and still immature in many aspects, in terms of both investors and risk management. This was most evident during the subprime mortgage crisis in 2007–2008 and the high-leverage capital allocation in 2015–2016, which resulted in panic and abnormal fluctuations of the stock market. Analyzing the evolution of networking connectivity during tranquil and turbulent periods, and exploring leading stocks and industry characteristics, may help investors optimize asset allocations; in addition, it may help policymakers identify information transmission paths, formulate regulatory policies, and even develop a bailout strategy.

Complex networks are widely applied to financial markets, such as stock connections, derivative price correlations, and investor sentiment \cite{chen2021dynamic,wen2019tail, ruan2018using, lee2019global, leur2017network, sun2020spillovers}; increasingly, scholars focus are focusing on the abnormal networking
structure during crisis periods \cite{lee2018state, zhao2016structure, hale2016crisis, loipersberger2018the,yu2020stock,yu2021earnings}, showing different national stock markets share similar patterns over several crises, while more heterogeneities exist between developed and developing countries. \citet{onnela2003dynamic}  employ the minimum spanning tree to construct the US equity market network from 1982 to 2000 and study the impact of Black Monday on the US market. \citet{zhao2016structure} use the correlation networks to analyze the impact of the subprime mortgage crisis on the US stock market and discuss the structural changes of the US stock network during the crisis. \citet{nobi2014effects} also build correlation networks, using the threshold method, to investigate the impact of the subprime mortgage crisis on the Korean stock market. \citet{coletti2017the} study the dynamic evolution of the Italian stock market during the 2008–2011 subprime mortgage crisis and the European debt crisis. Topological changes of the South African stock market during the subprime mortgage crisis are discussed by \citet{majapa2016topology}. Stock markets appear to attract much attention in the literature because of data availability and their profound influence on global finance. There is growing literature on the risk spillover between countries during crises. \citet{brechmann2013risk} study risk contagion in the European stock market networks during the global financial crisis. \citet{karmann2014volatility} discuss interactions and structural changes of Asian stock market networks during the Asian financial crisis. \citet{eom2010the} discover a significant increase in the information flow of national stock markets during the GFC. \citet{hardle2016tenet} use CoVaR and Pearson correlations to build stock networks and further analyze
systemic risks during periods of recession. \citet{kumar2012correlation} use Pearson correlations and the threshold method to construct a network consisting of 20 countries and explore topological changes of generated networks under different threshold values.

As China has an increasing influence on the global economy and finance, the dynamic evolution, risk spillover effects and abnormal volatility in the Chinese stock market is attracting more and more attention in the literature. \citet{tu2014cointegration} analyze the structural features and cointegration evolution of the Chinese stock market network during the 2008 subprime mortgage crisis. \citet{wang2018cross} and \citet{khoojine2019network} study the network characteristics and risk spillover of China’s stock market during the collapse of China’s stock market from 2015 to 2016 by considering network information, stock returns, and trading volumes. \citet{zhang2019the} further compare the differences in topological structures of the Chinese stock market in two recession periods of 2008 and 2015. Based on 83 constituents in Shanghai Shenzhen CSI 300 Index, their results confirm the differences of the SSE A-shares market between the 2008 and 2015 crises. \citet{huang2021an} find that the average clustering coefficient rises significantly during the financial downturns of the Chinese market. The microstructure and evolution features of the Chinese stock market are well discussed under the complex network framework. Most of the studies compare the differences of the SSE A-shares market in two abnormal volatility stages, yet lack any comparisons between tranquil and turbulent periods. Moreover, although some papers point out the aggregation effect and the abnormal increase of connectivity in stock networks during crises, few articles investigate underlying factors causing this phenomenon. There remain few studies on
the asymmetry network structures during the bubble accumulation periods and crises, especially in the emerging markets \cite{wu2019global,chen2021dynamic}.

%

Based on the external impact of the US subprime mortgage crisis on the Chinese stock market and the internal adjustment before and after the high-leverage capital allocation in 2015, this paper divides the long-term SSE A-shares market data into six stages by considering market rises and falls, the risk aggregation and release, and the fluctuation range of the stock market. Our work has four main contributions. First, an innovative threshold method is proposed to generate networks based on the degree distributions, $ 3 \sigma $ principle, the maximum strongly connected subgraph, and the iterative process. Second, the SSE A-shares network shows a more significant aggregation effect and stronger connectivity in turbulent periods than tranquil ones, and the connectivity and aggregation of the SSE A-shares network in 2015 are stronger than those in 2008. Third, analysis of key industries and node features driving the market suggests an asymmetric effect in ‘bull’ and ‘bear’ markets; leading nodes have the dominant status in bull markets, while leading nodes in bear markets have significant fat-tailed distributions. Finally, the multivariate secondary assignment method is used to analyze factors affecting the node degree rank. Empirical results show that the degrees of those leading stocks in bear markets are significantly and positively correlated with their market values, and off-site funding plays a more critical role than margin trading in the massive declines during the 2015 high-leverage capital allocation.


\section{Data and Methodology}

\subsection{Data Description and Stage Division}

\begin{figure}
	\begin{center}
		\includegraphics[width=\linewidth]{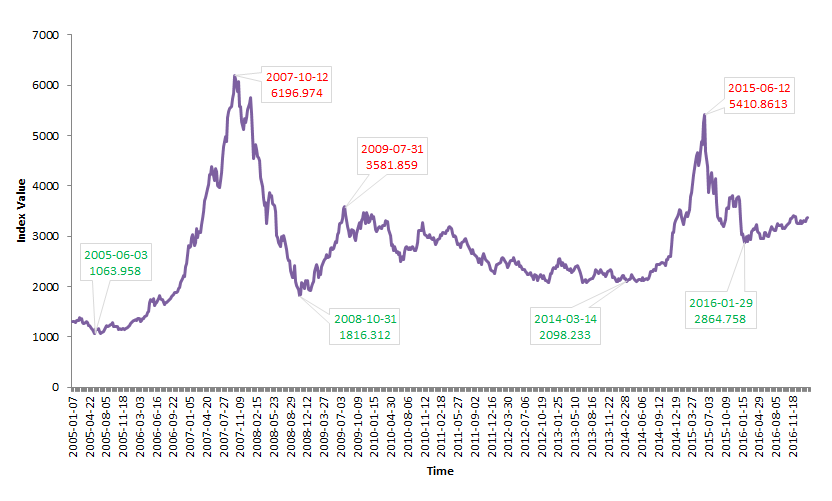}
		\caption{\label{Fig SSE Index Trend} SSE composite index trend.}	
	\end{center}	
\end{figure}
\begin{figure}
\begin{center}
	\includegraphics[width=\linewidth]{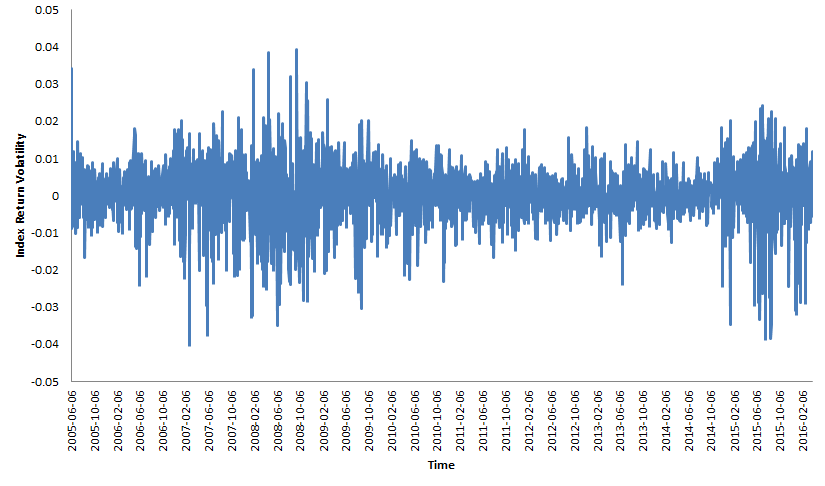}
	\caption{\label{Fig SSE Index Return Volitility} SSE composite index return volitility.}	
\end{center}			
\end{figure}

This paper uses the SSE A-shares market’s weekly closing prices from 2005 to 2016. Using important national and international financial events and policies, the data is divided into six stages according to whether they represent bull markets (periods of growth) and bear markets (periods of recession), summarized in Table~\ref{Tab Period Division from 2005 to 2016} below. The trend and return volatility of SSE composite index are presented in Figures~\ref{Fig SSE Index Trend} and \ref{Fig SSE Index Return Volitility} respectively.

\begin{table}
	\centering
	\caption{Stage division for the SSE A-share market.}
	\label{Tab Period Division from 2005 to 2016}
	\begin{tabular}{ccl}
		\toprule
		\multicolumn{1}{l}{Stage} & \multicolumn{1}{l}{Market feature} & Reasons                                                                                                                                                           \\ \hline
		2005/06/03--2007/10/12    & BULL 1                             & The stock reform stimulates market                                                                                                                      \\
		&                                    & liquidity. Domestic trade surplus    \\
		& & and RMB (Renminbi) appreciation                \\
		& & lead to risk aggregation.                                                                  \\ \hline
		2007/10/12--2008/10/31    & BEAR 1                             & The subprime mortgage crisis and   \\
		& & other external factors lead to bubbles   \\
		& & in industrials and real estate.                                                                                                                                                            \\ \hline
		2008/10/31--2009/07/31    & BULL 2                             & Rescue policies from the government \\
		& & promote econmic recovery.                                                                                                                                                                                                                                                               \\ \hline
		2009/07/31--2014/03/14    & BEAR 2                             & The suspension of domestic IPO   \\
		& & and European debt crisis.                                                                                                                                                                                                                                                                         \\ \hline
		2014/03/14--2015/06/12    & BULL 3                             & Deepen reforms of state-owned   \\
		& & enterprises, arising financial leverages  \\
		& & increase the market and the rapid   \\
		& & development of marginal trading and \\
		& &  off-side funding.                                                                                                                                                         \\ \hline
		2015/06/12--2016/01/29    & BEAR 3                             & Deleveraging process and supervision                                                                                         \\
		& & on shadowbanks and off-side funding.                                                                                                                                                                  \\ \bottomrule
	\end{tabular}
\end{table}

As shown in Figures~\ref{Fig SSE Index Trend} and \ref{Fig SSE Index Return Volitility}, the significant structural differences between bull and bear periods are evident in the SSE A-shares market, suggesting the SSE A-shares market is closely related to macroeconomic policies. Moreover, the first and third stages (BULL 1 and BULL 2) witness a surge in share prices, and these prices experience a rapid fall in the second and further stages (BEAR 1 and BEAR 2). Such sharp amplitudes in stock prices undermine the stability of the financial market and contribute to system risks, highlighting the immaturity of capital markets in emerging countries. Given policy factors, equity price fluctuations, and return volatilities, the long-term cycle in the SSE A-shares market is divided into three stages – BULL 1 and BEAR 1, BULL 2 and BEAR 2, and BULL 3 and BEAR 3. BULL 1 and BEAR 1, and BULL 3 and BEAR 3, belong to the abnormal volatility stage, where the volatility grows rapidly in these four periods. Specifically, BULL 1 and BULL 3 show a surging trend due to the irrational rise of the index in the short-term, accumulating bubbles and risks.  BEAR 1 and BEAR 3 witness dramatic declines, and thousands of stocks in the market repeatedly hit limit down, manifesting as a collapsed state. In contrast, BULL 2 and BEAR 2 belong to an exponential oscillation stage and have relatively stable volatilities, regarded as relatively tranquil periods. By comparing the first (BULL 1 and BEAR 1) and third (BULL 3 and BEAR 3) cycle with the second cycle (BULL 2 and BEAR 2), we investigate the structural differences between turbulent and tranquil periods. We also compare the first and third cycles to study the external impact of the subprime mortgage crisis on the SSE A-shares market and the diverse effects of internal ``leverage and deleveraging'' liquidity shift on the market.


\subsection{Data}


The SSE A-shares backward closing prices from 2005 to 2016 are downloaded from the Wind Database. Given the long study period, some stocks are removed from the sample for the following reasons:
\begin{enumerate}
	\item Missing values. Due to the late listing and suspension of trading, some of these stocks were removed in advance to compare stock market networks in different stages.
	
	\item Stocks prefixed with ``ST'' or ``*ST''. Designated as Special Treatment (ST) by the stock exchanges to warn investors, these stocks face delisting risks and show distinct patterns from normal stocks.
	
	\item Stocks who maintain a zero return over a long period, either due to long-term suspension or other reasons, were excluded in this paper to prevent misleading information.
\end{enumerate}

For the stock $ i $, its log return is calculated by
\begin{equation}
	r_{it} = \ln P_{it} -  \ln P_{i,t-1},
	\label{Eq Log Returns}
\end{equation}
where $ P_{it} $ is the price of stock $ i $ at time $ t $. As a vital measurement to qualify the relationship between two objections, the Pearson correlation is widely used in the construction of stock networks \cite{wang2018correlation, ruan2018using, zhang2015dynamic, lee2019global} to reveal direct co-movements among equities (e.g., stocks within the same industrial sector are subject to the same movements) and to discover potential stock relationships such as holding companies and asymmetric information. We also consider the Granger linear and nonlinear causality test to depict the before-after relationships in the SSE A-shares market and compare causality networks with the dynamic correlation network.
%
Then, we based on the returns of stocks $ i $ and $ j $ to compute the sample Pearson correlation
\begin{equation}
	\widehat{\rho}_{ij} = \frac{\sum_{t=1}^{T} \left(r_{it} - \bar{r}_i \right) \left(r_{jt} - \bar{r}_j \right) }{\sqrt{\sum_{t=1}^{T} \left(r_{it} - \bar{r}_i \right)^2} \sqrt{\sum_{t=1}^{T} \left(r_{jt} - \bar{r}_j \right)^2}},
	\label{Eq Pearson Correlations}
\end{equation}
where $ \bar{r}_i = \frac{1}{T} \sum_{t=1}^{T} r_{it} $ is the sample mean of stock $ i $'s returns. Absolute values of Pearson correlations in Equation~(\ref{Eq Pearson Correlations}) are used to measure the relationship between $ i $ and $ j $ and further determine the threshold value over different periods.


\subsection{Determining the Threshold Value for Network Construction}


The adjacency matrix is the basis for studying complex networks. As one of the mainstream approaches to generate networks, the threshold method only selects connections within a given range determined by thresholds. Since different threshold values give rise to distinct networking structures and topological properties, determining appropriate thresholds is essential – especially for dynamic networks. This article improves the classic threshold method in three ways. First, we adopt a uniform threshold to compare network structures
over different stages. Second, the threshold is set to filter trivial relationships by highlighting the status of HUB stocks in the market. Finally, we include the maximum strongly connected subgraphs step in our method to avoid losing too much information. The specific approach is as follows.

\paragraph{Step 1:  Use the $ 3 \sigma $ principle to determine the threshold range shared by all stages}
 This can be found using the probability distributions of Pearson correlations in the six stages the threshold is highly likely to be located in $ \left[\mu - 3 \sigma, \mu + 3 \sigma\right] $, where $ \mu $ and $ \sigma $ are the sample mean and sample standard deviation of sample Pearson correlations respectively. If the threshold value is set to be less than $ \mu $, more than $ 50 \% $ of connections will be considered in the generated network, which inevitably involves too many trivial connections. Therefore, $ \left[\mu, \mu + 3 \sigma\right] $ is viewed as the reasonable interval and Table~\ref{Tab Basic statistics of sample Pearson correlations} presents basic statistics of the sample Pearson correlations in each stage. Given by the structural changes of the SSE A-shares network in different stages, the shared range of $ \left[\mu - 3 \sigma, \mu + 3 \sigma\right] $ in six stages is $ \left[0.5478, 0.6640\right] $, implying a reasonable range of the uniform threshold for all periods.

\begin{table}
	\begin{spacing}{1.5}
		
		\centering
		\caption{Basic statistics of sample Pearson correlations over six stages.}
		\label{Tab Basic statistics of sample Pearson correlations}

		\begin{tabular}{cccccc}
			\toprule
			Stage  & Sample mean $ \mu $ & Sample SD $ \sigma $ & $ \mu - 3 \sigma $ & $ \mu + 3 \sigma $ & $ \left[\mu, \mu + 3 \sigma \right] $ \\ \hline
			BULL 1 & 0.3629              & 0.1383                               & -0.052  & 0.7778 & {[}0.3629,0.7778{]} \\
			BEAR 1 & 0.5478              & 0.145                                & 0.1127  & 0.983  & {[}0.5478,0.9830{]} \\
			BULL 2 & 0.542               & 0.1559                               & 0.0744  & 1.0095 & {[}0.5420,1.0095{]} \\
			BEAR 2 & 0.3764              & 0.0959                               & 0.0889  & 0.664  & {[}0.3764,0.6640{]} \\
			BULL 3 & 0.2633              & 0.1482                               & -0.1813 & 0.7078 & {[}0.2633,0.7078{]} \\
			BEAR 3 & 0.5398              & 0.2268                               & -0.1405 & 1.2202 & {[}0.5398,1.2202{]} \\ \bottomrule
		\end{tabular}
		
	\end{spacing}
\end{table}

\paragraph{Step 2: Determine the threshold range by maximum strongly connected subgraphs}

\begin{table}
	\centering
	\caption{Node numbers of maximum strongly connected subgraphs of the SSE A-shares network in six stages.}
	\label{Tab Node numbers of maximum strongly connected subgraphs}
	\begin{tabular}{crrrrrr}
		\toprule
		Threshold & \multicolumn{1}{c}{BULL 1} & \multicolumn{1}{c}{BEAR 1} & \multicolumn{1}{c}{BULL 2} & \multicolumn{1}{c}{BEAR 2} & \multicolumn{1}{c}{BULL 3} & \multicolumn{1}{c}{BEAR 3} \\ \hline
		0.2       & 605                        & 605                        & 605                        & 605                        & 605                        & 605                        \\
		0.3       & 605                        & 605                        & 605                        & 605                        & 605                        & 605                        \\
		0.4       & 605                        & 605                        & 605                        & 602                        & 602                        & 605                        \\
		0.5       & 585                        & 604                        & 600                        & 569                        & 571                        & 603                        \\
		0.6       & 428                        & 589                        & 595                        & 338                        & 436                        & 591                        \\
		0.7       & 132                        & 543                        & 558                        & 50                         & 138                        & 551                        \\
		0.8       & 2                          & 325                        & 387                        & 4                          & 6                          & 462                        \\
		0.9       & 1                          & 2                          & 4                          & 1                          & 1                          & 206                        \\ \bottomrule
	\end{tabular}
\end{table}

Table~\ref{Tab Node numbers of maximum strongly connected subgraphs} presents node numbers of maximum strongly connected subgraphs in the SSE A-shares network under different threshold values. As suggested in Table \ref{Tab Node numbers of maximum strongly connected subgraphs}, the node number of the maximum strongly connected subgraph in each stage changes significantly when the threshold value is located between $ 0.6 $ and $ 0.7 $. In particular, the maximum strongly connected subgraph involves too many nodes when the threshold is less than $ 0.6 $, which means that the built network may contain too much redundant information. By contrast, when the threshold is higher than $ 0.7 $, the networking structure becomes too sparse to reflect all key relationships in the stock market. Given the $ 3 \sigma $ principle, we further narrow the threshold range to $ [0.6, 0.6640] $. This step further narrows the threshold range to maintain the key structures in the financial market (e.g., HUB stocks) and protect these micro subgraphs from disruption of trivial relationships.

\paragraph{Step 3: Determine the final uniform threshold}
Within the new threshold interval $ [0.6,0.6640] $, the maximum strongly connected subgraphs in BEAR 1, BULL 2, and BEAR 3 share a similar node number, implying the relatively stable networking structure. We further discuss the relationships between thresholds and the maximum strongly connected subgraphs in Table~\ref{Tab Relationships between thresholds and maximum strongly connected subgraphs} to decide the final threshold. As suggested in Table~\ref{Tab Relationships between thresholds and maximum strongly connected subgraphs}, the node number in maximum strongly connected subgraphs over all stages remains relatively stable as the threshold increases from $ 0.63 $ to $ 0.65 $, implying the stability of generated networks. Within the range $ [0.63, 0.65] $, we consider a smaller interval length $ d = 0.01 $ and find the interval $ [0.64, 0.65] $ within which the node number of the  maximum strongly connected subgraphs is most stable. Iterate this procedure to narrow the threshold range until $ d = 0.0001 $, and the uniform threshold is set to $ \theta_0 = 0.6456 $.

\begin{table}
\centering
\caption{Relationships between thresholds and node numbers of the maximum strongly connected subgraphs in BULL 1, BEAR 1 and BULL 3.}
\label{Tab Relationships between thresholds and maximum strongly connected subgraphs}
\begin{tabular}{crrrrrr}
	\toprule
	\multirow{2}{*}{Threshold} & \multicolumn{2}{c}{BULL 1}                                             & \multicolumn{2}{c}{BEAR 2}                                              & \multicolumn{2}{c}{BULL 3}                                              \\ \cline{2-7} 
	& \multicolumn{1}{c}{Node} & \multicolumn{1}{c}{Reduction} & \multicolumn{1}{c}{Node} & \multicolumn{1}{c}{Reduction} & \multicolumn{1}{c}{Node} & \multicolumn{1}{c}{Reduction} \\
	& \multicolumn{1}{c}{number} & \multicolumn{1}{c}{number} & \multicolumn{1}{c}{number} & \multicolumn{1}{c}{number} & \multicolumn{1}{c}{number} & \multicolumn{1}{c}{number} \\ \hline
	0.60                       & 428                             & -                                    & 338                             & -                                    & 436                             & -                                    \\
	0.61                       & 395                             & -33                                  & 281                             & -57                                  & 408                             & -28                                  \\
	0.62                       & 372                             & -23                                  & 237                             & -44                                  & 366                             & -42                                  \\
	0.63                       & 340                             & -32                                  & 202                             & -35                                  & 344                             & -22                                  \\
	0.64                       & 311                             & -29                                  & 166                             & -36                                  & 316                             & -28                                  \\
	0.65                       & 294                             & -17                                  & 144                             & -22                                  & 282                             & -34                                  \\
	0.66                       & 253                             & -41                                  & 92                              & -52                                  & 254                             & -28                                  \\ \bottomrule
\end{tabular}

\end{table}


\subsection{Network Construction}

Let $ \bm{A} $ be the adjacency matrix of the generated network under the optimal threshold $ \theta_0 $ and $ a_{ij} $ be the $ (i, j) $th element of $ \bm{A} $. Then, $ a_{ij} $ is defined as
\begin{equation}
	a_{ij} = \left\lbrace \begin{array}{cc}
		1,   &   \left| \hat{\rho}_{ij} \right| \geq \theta_0,   \\
		0,   &   \left| \hat{\rho}_{ij} \right|  <   \theta_0.
	\end{array} \right.
	\label{Eq Adjacency Matrix}
\end{equation}
Here, $ \theta_0 $ is set to be $ 0.6456 $.


\subsection{Network Topological Properties}

We consider the following network topological properties to investigate the evolution of the SSE A-shares network.

\begin{description}
	
	\item[Average shortest path length] $ L = \frac{2}{N(N-1)} \sum_{j \neq i} d_{ij} $, where $ i $ and $ j $ are two stocks (nodes) in the SSE A-shares network and $ d_{ij} $ is the shortest path between nodes $ i $ and $ j $. A smaller length means faster information or risk transmission in the network.
	
	\item[Clustering coefficient] $ C_i = \frac{2 n_i}{k_i (k_i - 1)} $, where $ k_i $ is the number of nodes directly connecting to node $ i $ and $ n_i $ is the number of edges between $ k_i $ neighbours of node $ i $. A higher value implies better network connectivity.
	
	\item[Network diameter] $ \mathrm{Diameter} = \max_{1 \leq i,j \leq N} d_{ij} $. A smaller value implies faster information or risk transmission speed.
	
	\item[Network density] $ \mathrm{Density} = \frac{\sum_{i,j} a_{ij}}{N (N-1)} $, where $ a_{ij} $ is defined in Equation (\ref{Eq Adjacency Matrix}). A higher density implies a closer relationships between nodes.
	
	\item[Relative degree centrality] $ C_{RD} (i) = \frac{k_i}{N-1} $. The high relative degree centrality implies an important influence from the corresponding node on the network. 
	
	\item[Relative betweenness centrality] $ C_{RB} (i) = \frac{2}{(N-1)(N-2)} \sum_{j < k} \frac{g_{jk} (i)}{g_{jk}} $, where $ g_{jk} (i) $ is the number of shortest paths connecting nodes $ j $ and $ k $ and passing through node $ i $. This indicator measures the ``bridge'' role of node $ i $ in the network.
	
	\item[Relative closeness centrality] $ C_{RC} (i) = \frac{N-1}{\sum_{j=1}^{N} d_{ij}} $ that measures how close node $ i $ is to all other nodes in the network. The high value of relative closeness centrality implies close connections between node $ i $ and other nodes.
	
	\item[Degree centralization] $ C_D = \frac{\sum_{i=1}^{N} \left( C_{RD} (\max) - C_{RD}(i) \right)}{\max \left[ \sum_{i=1}^{N} \left( C_{RD} (\max) - C_{RD}(i) \right) \right]} $, where the numerator is the sum of differences between the maximum degree centrality $ C_{RD} (\max) = \max_{1 \leq i \leq N} C_{RD}(i) $ and the degree centrality of each node $ C_{RD} (i) $, and the denominator is the maximum value of the numerator. This indicator describes the centrality of the whole network.
	
	\item[Betweenness centralization] $ C_B = \frac{1}{N-1} \sum_{i=1}^{N} \left( C_{RB} (\max) - C_{RB}(i) \right) $, where
	the numerator theoretically represents the sum of the difference between the maximum intermediate centrality and the intermediate centrality of each node, and the denominator represents the maximum of the sum of the differences. This indicator describes the degree to which the network relies excessively on a node to transfer relations.

	\item[Closeness centralization] $ C_C = \frac{2(N-3)}{(N-1)(N-2)} \sum_{i=1}^{N} \left( C_{RC} (\max) - C_{RC}(i) \right) $, which describes the centralised trend in the network.
	
	\item[Scale-free networks] Denote $ P(k) $ by the probability of a node with degree $ k $ in a given network. The scale-free network satisfies $ P(k) \propto k ^{- \lambda} $, where $ \lambda > 0 $ is the scale-free parameter. The node distribution of a scale-free network follows the power law, which means that most nodes have low degrees while a few nodes have high degrees. Referred to as HUB, these high-degree nodes play critical roles in risk transmission because of their massive connections with other nodes.

	\item[Heterogeneity] Denote $ D_i $ by the degree of node $ i $. The heterogeneity is defined as $$ \frac{\sum_{i=1}^{N} D_i^2}{\left(\sum_{i=1}^{N} D_i\right)^2}, $$ which measures the difference of node statuses in a given network. The low heterogeneity implies there is little difference among nodes.
	
\end{description}


\section{Empirical Results of SSE A-Shares Dynamic Correlation Network}

\subsection{Analysis of Dynamic Correlation Network in SSE A shares}

\begin{figure}
\begin{center}
	\subfloat[\label{fig:location1}BULL 1]{\includegraphics[scale=0.25]{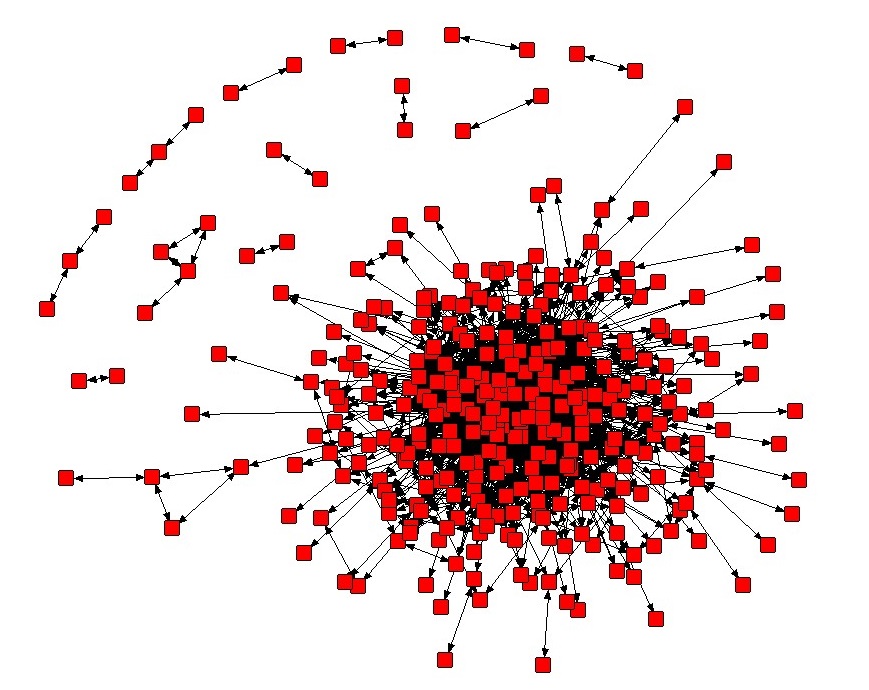}}
	\subfloat[\label{fig:location2}BEAR 1]{\includegraphics[scale=0.25]{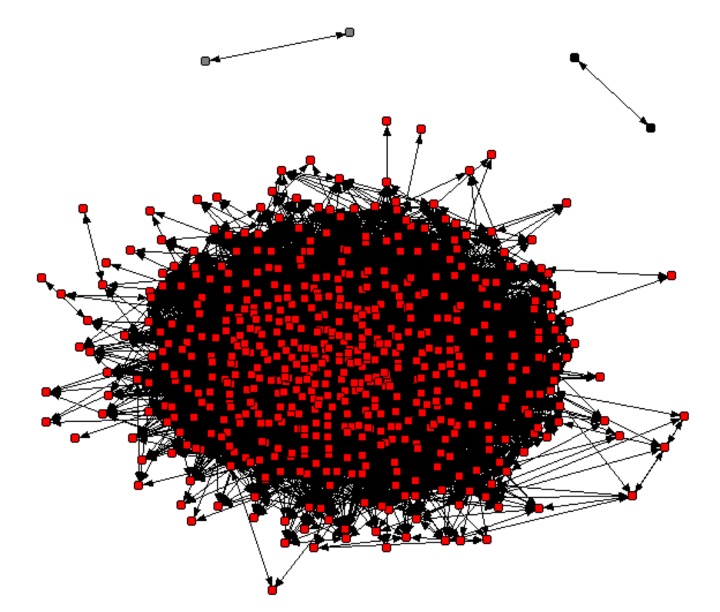}}\\
	\subfloat[\label{fig:location3}BULL 2]{\includegraphics[scale=0.25]{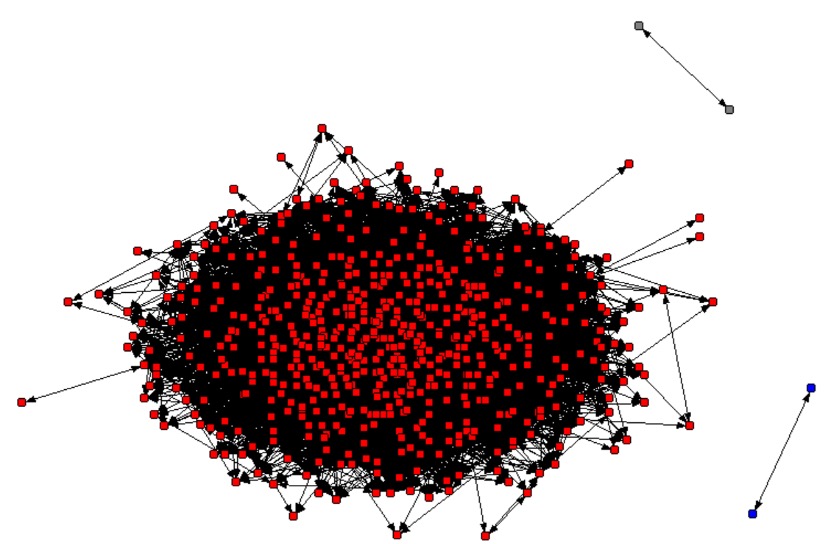}}
	\subfloat[\label{fig:location4}BEAR 2]{\includegraphics[scale=0.25]{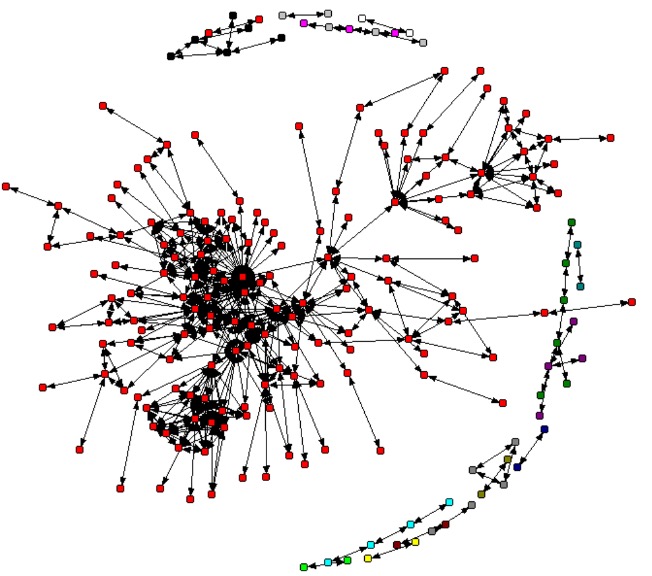}}\\
	\subfloat[\label{fig:location5}BULL 3]{\includegraphics[scale=0.25]{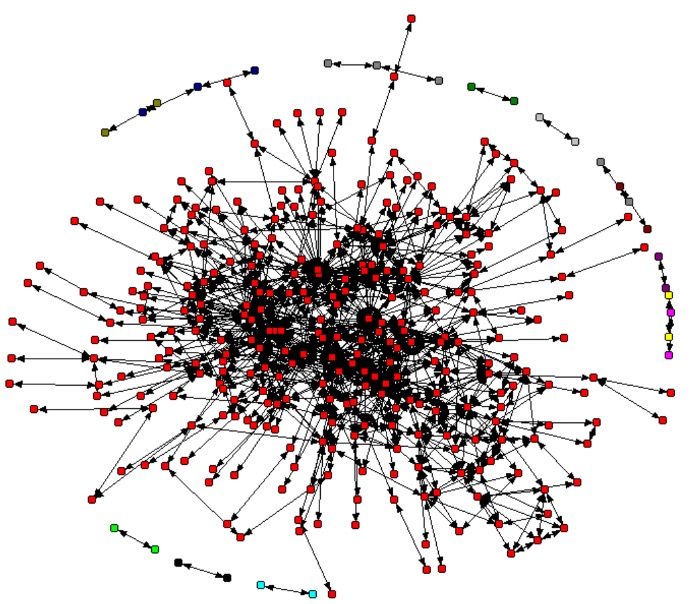}}
	\subfloat[\label{fig:location6}BEAR 3]{\includegraphics[scale=0.25]{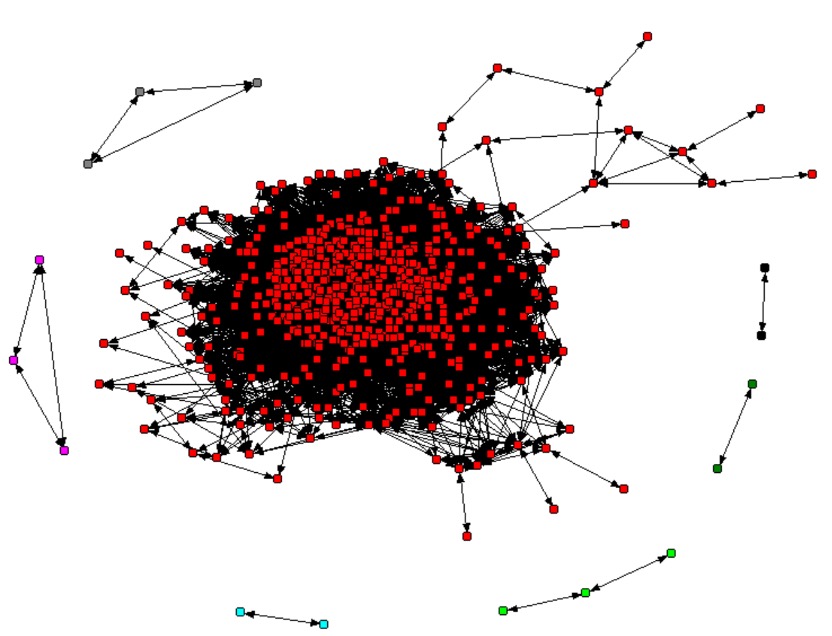}}
	\caption{\label{Dynamic correlation network composition of A Shares in Shanghai Stock Exchange}Dynamic correlation network composition of SSE A-Shares.}
\end{center}	
\end{figure}

Figure~\ref{Dynamic correlation network composition of A Shares in Shanghai Stock Exchange} depicts the composition diagram of the dynamic correlation network of SSE A-shares in each stage. There is a connection between the nodes belonging to the same component, but there is no connection between the nodes of different components. The more components, the lower the correlation degree of stock nodes in the market.

As can be seen from Figure~\ref{Dynamic correlation network composition of A Shares in Shanghai Stock Exchange}, most of the nodes in each phase belong to the same component The phases with the strongest connectivity and the fewest marginal nodes, which are almost the same component, are BEAR 1 and BEAR 3; they represent the down phases in abnormal stock market volatility. The phase with the most marginal components, BEAR 2, is the oscillating bear phase. The network compositions show that while the tranquil and turbulent periods differ greatly in terms of macro connectivity, there is also asymmetry between bulls and bears, and the connectivity of bears is stronger than that of bulls. We also calculate the topological properties of the network in different phases, shown in Table~\ref{Results of Dynamic Associated Network Topology of Shanghai A Stock Exchange}.

\begin{table}
\centering
\caption{Dynamic correlation network topologies of the SSE A-shares market.}
\label{Results of Dynamic Associated Network Topology of Shanghai A Stock Exchange}
\begin{tabular}{lcccc}
	\toprule
	Stage & Average path length & Cluster coefficient & Diameter & Density \\ \midrule
	BULL 1 & 2.788               & 0.459               & 8        & 0.0115  \\
	BEAR 1 & 1.794               & 0.721               & 5        & 0.2710  \\
	BULL 2 & 1.790               & 0.702               & 5        & 0.2765  \\
	BEAR 2 & 3.818               & 0.550               & 9        & 0.0024  \\
	BULL 3 & 3.720               & 0.381               & 10       & 0.0049  \\
	BEAR 3 & 1.730               & 0.802               & 7        & 0.3976  \\ \bottomrule
\end{tabular}	
\end{table}

The results in Table~\ref{Results of Dynamic Associated Network Topology of Shanghai A Stock Exchange} show that:
\begin{enumerate}

\item Comparing the two abnormal fluctuation periods shows that the small-world effect is stronger in BEAR 3 than in BEAR 1. In the early stage of the market, high leverage capital allocation was caused by risk aggregation in BEAR 3. Then regulatory tightening and other factors such as rapid deleveraging, led to a drop in the stock market. The rapid enhancement of network connectivity further strengthened the risk contagion in the bear period and increased the magnitude of the loss. In BEAR 1, due to the global market fluctuations caused by the subprime mortgage crisis, as well as the relatively weak capital flows in China's financial and external markets in 2008, the abnormal volatility and rapid changes in connectivity posed by external shocks in 2008 were significantly more vulnerable than those in 2015.

\item The comparison of the bull market and bear markets during abnormal fluctuation periods shows an asymmetric effect between a fast falling bear market and a rising bull market in an abnormal fluctuation period, and the small-world effect of the bear market is significantly stronger than that of the bull market. First, the average path length of the stock market is lower than that of the bull market in the two abnormal fluctuating bear markets in BEAR 1 and BEAR 3; and, the average cluster coefficient is much higher than that of the bull market, indicating that the aggregation between the A-shares market is stronger in the bear market than the bull market period. Second, the diameter of the bear market is smaller than the bull market, and the maximum distance between nodes is less than $ 7 $, indicating that the distribution of nodes in the bear market stage is very compact. Third, the density difference between the bull and bear market periods is also substantial, with the density in the bear phase being much larger than the network density value in the bull phase, and the same down effect in the bear market is stronger than the same up effect in the bull market.

\item Comparing the tranquil and turbulent periods, we find that the network connectivity and risk contagion in the abnormal fluctuation periods are stronger than those in the normal fluctuation periods. The bear market in the general fluctuation period has the longest average path of the network with a weak small-world effect; this is in sharp contrast to the bear market in the abnormal fluctuation stage. Although BULL 2 and BEAR 2 are defined as a fluctuated bull market and shock bear market, respectively, the network topology of the two shock markets is unstable because of the uncertainty of market price. The market in BULL 2 was influenced by the 4 trillion RMB stimuli from the Chinese government. However, after the subprime mortgage crisis and the subsequent European debt crisis, the external impact was still powerful; despite the slightly falling overall index, the market showed a long period of ups and downs. Hence, regardless of the bull market and bear market, the topological nature of the relatively abnormal volatility stage is not distinct from the nature of the abnormal volatility period of bull and bear market characteristics.	

\end{enumerate}

After the analysis of the basic network topological properties, we further compare and check the similarities and differences of the network structures by using linear and nonlinear causal networks. Our findings show that the linear causal network maintains the mutual close relationship between the nodes. In constrast, the nonlinear causal network is very sparse, and there is no nonlinear causal relationship between most of the nodes; as a consequence, the topological properties of many stages cannot be compared and therefore these results mainly provide a comparative analysis of the linear causal network and the multi-stage dynamic threshold correlation network, as shown in Table~\ref{Comparison of network central potentials with different linear relations}.

\begin{table} 
\centering
\caption{Comparison of network central potentials with different linear relations}
\label{Comparison of network central potentials with different linear relations}
\begin{tabular}{lccc}
	\toprule
	Stage & \multicolumn{3}{c}{Pearson threshold network}  \\ \midrule
	& Relative       & Relative       & Relative       \\ 
	& degree         & between        & closeness      \\
	& centralization & centralization & centralization \\ \midrule
	BULL 1                  & 14.62\%        & 1.86\%         & 43.04\%        \\
	BEAR 1                  &  2.58\%        & 0.83\%         & 50.26\%        \\
	BULL 2                  & 39.21\%        & 90.00\%        & 44.60\%        \\
	BEAR 2                  &  5.90\%        & 2.44\%         & 23.42\%        \\
	BULL 3                  &  6.98\%        & 4.17\%         & 34.33\%        \\
	BEAR 3                  & 33.70\%        & 2.61\%         & 41.55\%        \\ \bottomrule
\end{tabular}	
\end{table}

\begin{figure}
\begin{center}
	\subfloat[\label{fig:location111}BULL 1]{\includegraphics[scale=0.49]{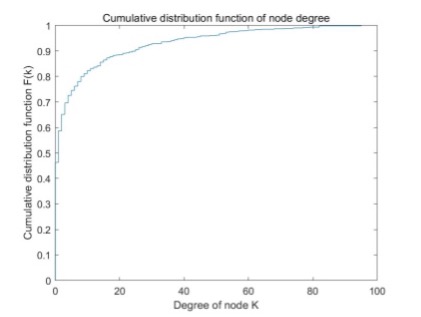}}
	\subfloat[\label{fig:location211}BEAR 1]{\includegraphics[scale=0.49]{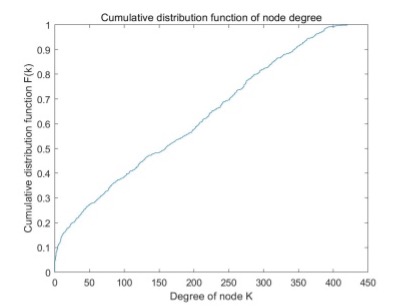}}\\
	\subfloat[\label{fig:location311}BULL 2]{\includegraphics[scale=0.49]{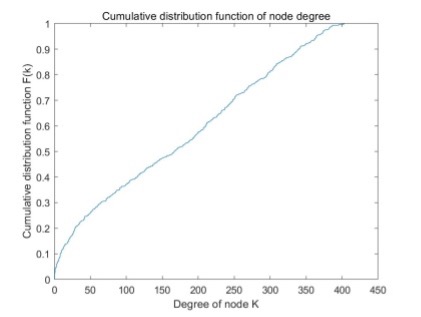}}
	\subfloat[\label{fig:location411}BEAR 2]{\includegraphics[scale=0.49]{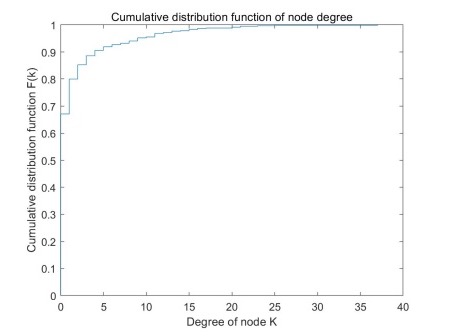}}\\
	\subfloat[\label{fig:location511}BULL 3]{\includegraphics[scale=0.49]{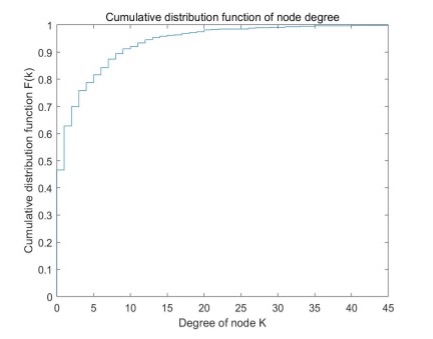}}
	\subfloat[\label{fig:location611}BEAR 3]{\includegraphics[scale=0.49]{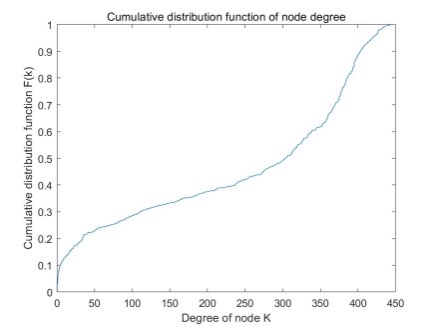}}
	\caption{\label{CDF of node degrees in stages}CDF of node degrees in stages.}
\end{center}		
\end{figure}

\begin{figure}
\begin{center}
	\subfloat[\label{fig:location11}BULL 1]{\includegraphics[scale=0.19]{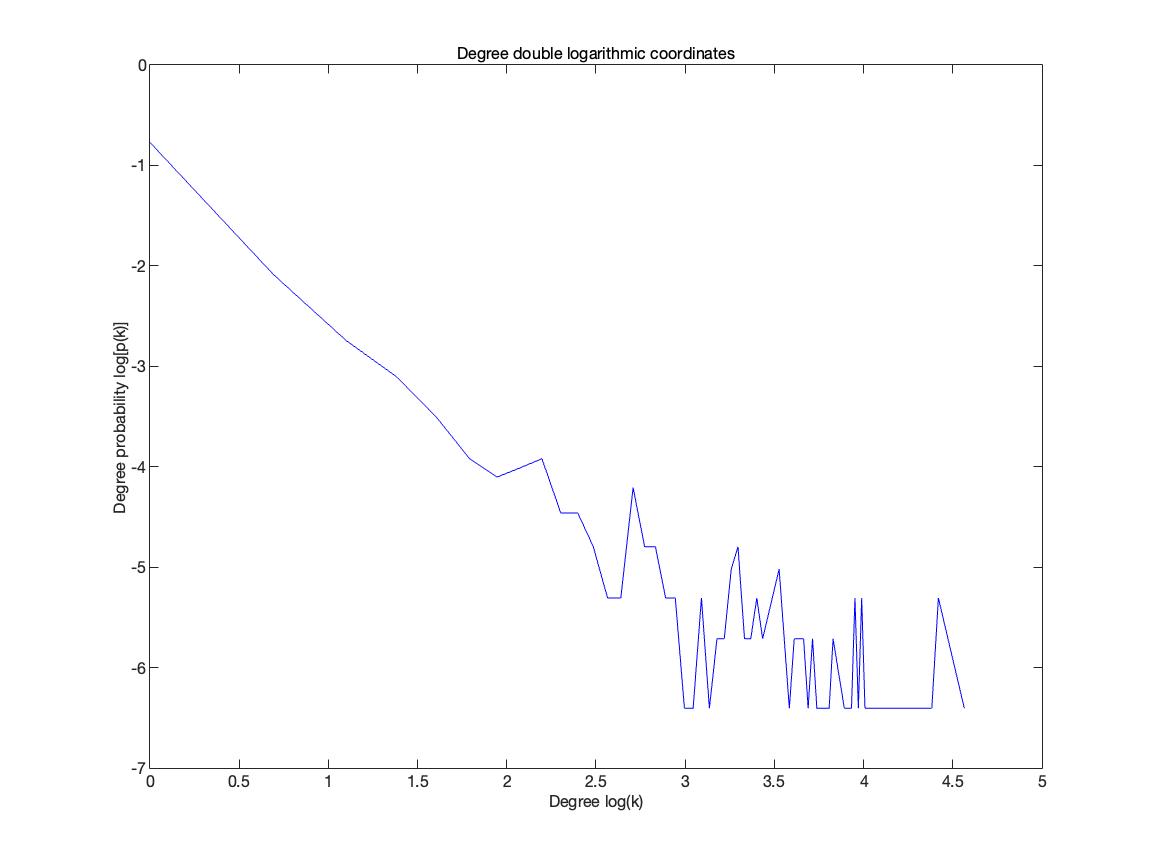}}
	\subfloat[\label{fig:location21}BEAR 1]{\includegraphics[scale=0.19]{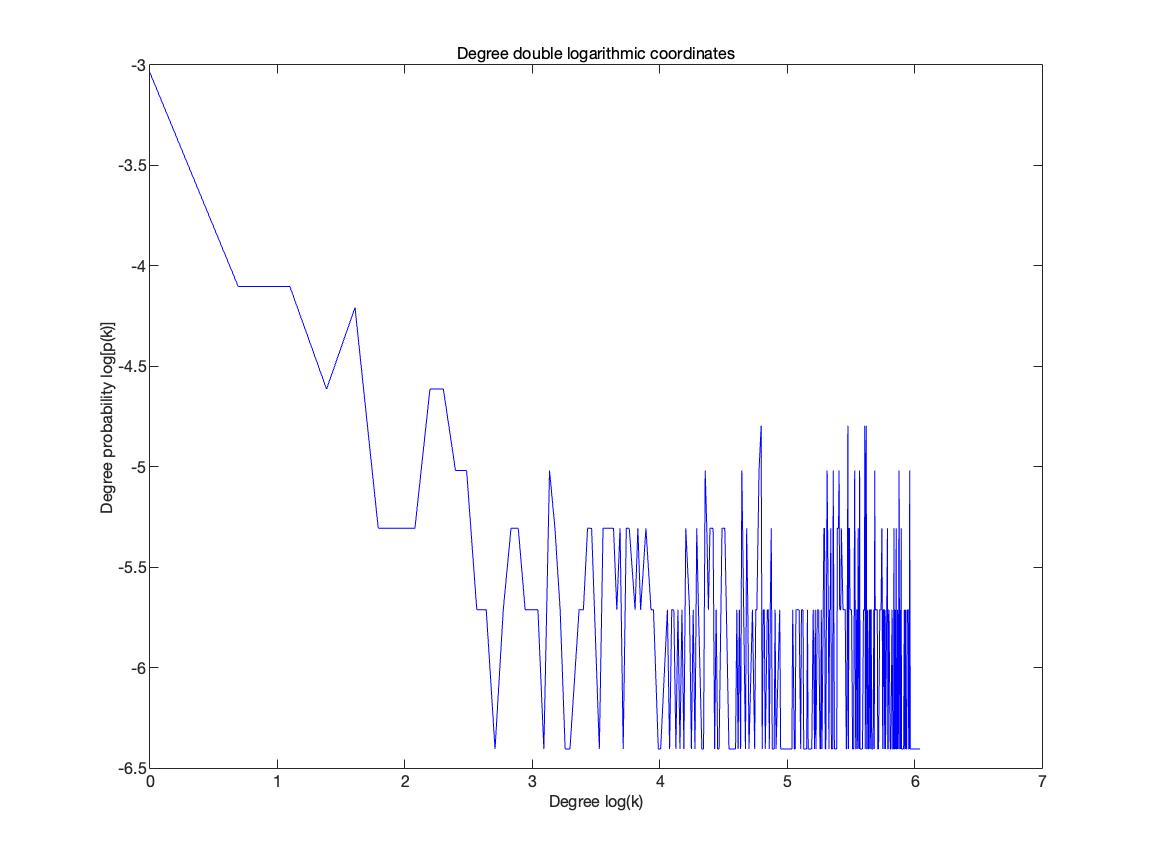}}\\
	\subfloat[\label{fig:location31}BULL 2]{\includegraphics[scale=0.19]{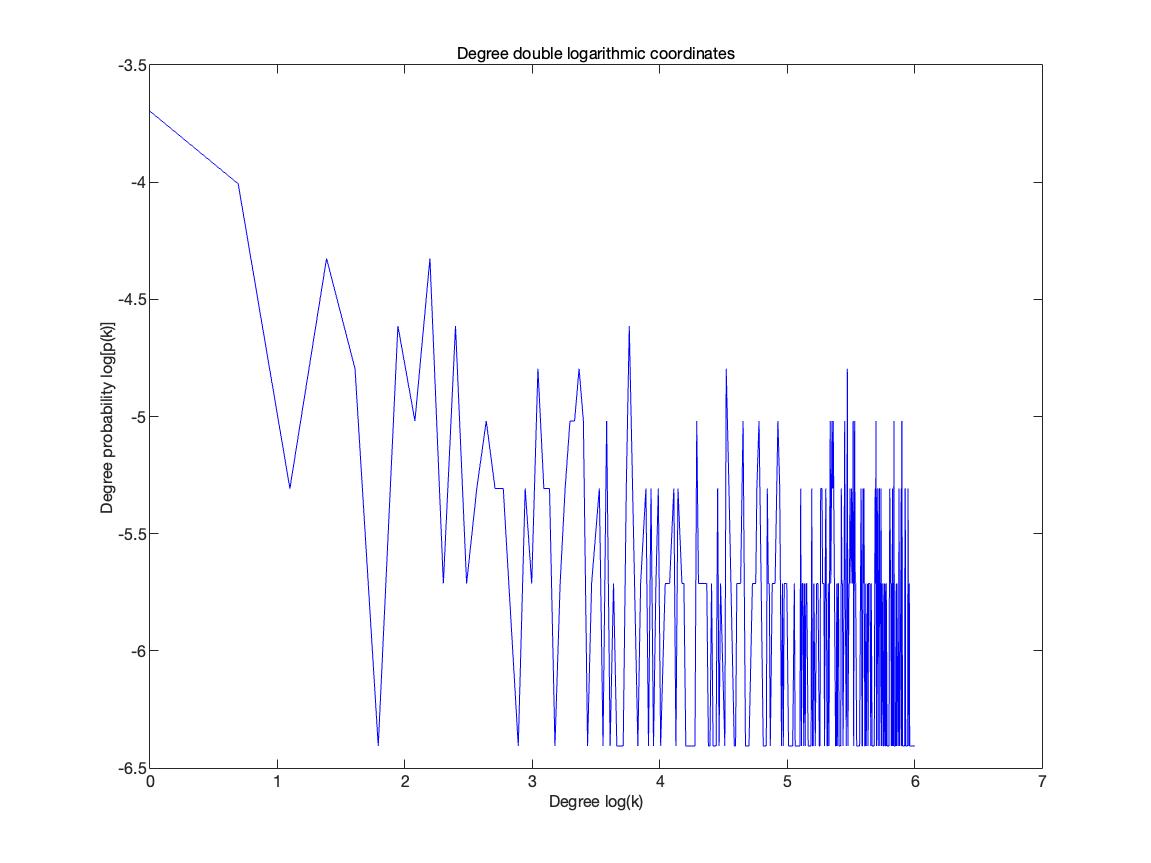}}
	\subfloat[\label{fig:location41}BEAR 2]{\includegraphics[scale=0.19]{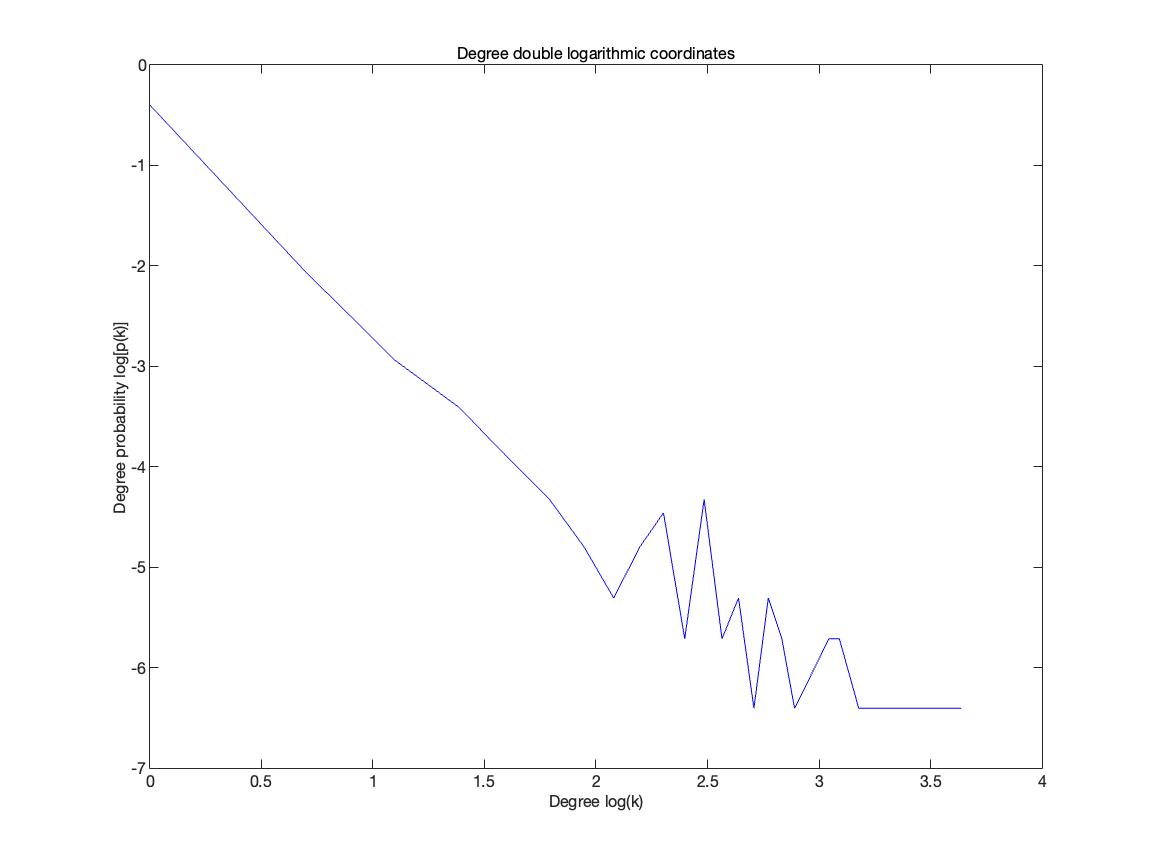}}\\
	\subfloat[\label{fig:location51}BULL 3]{\includegraphics[scale=0.19]{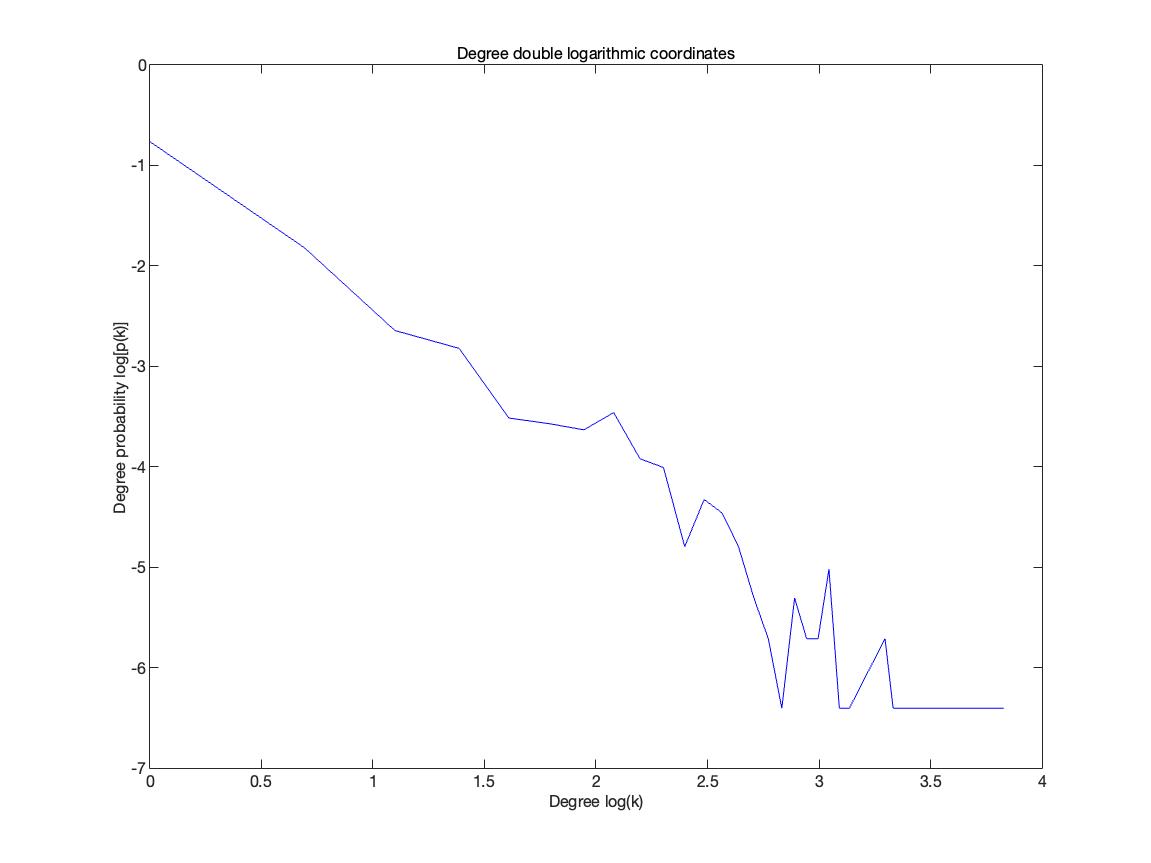}}
	\subfloat[\label{fig:location61}BEAR 3]{\includegraphics[scale=0.19]{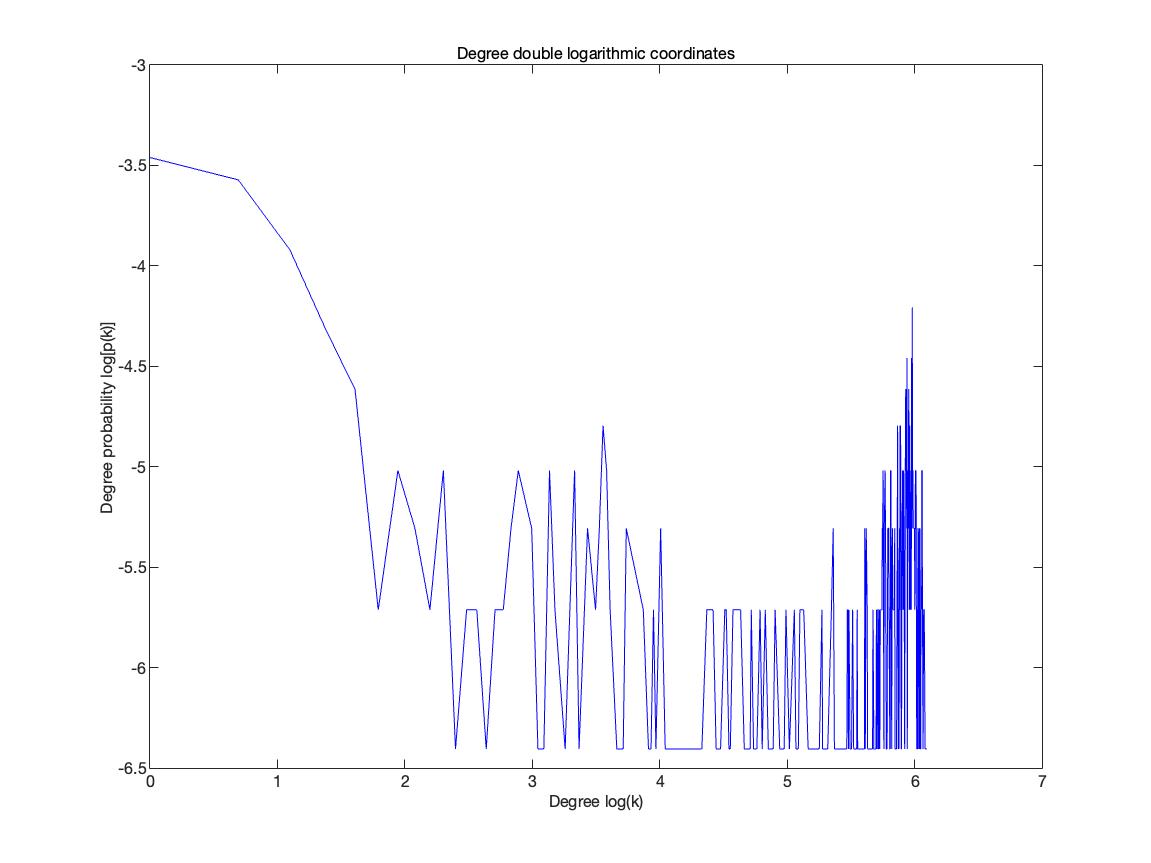}}
	\caption{\label{Double logarithmic distribution of node degrees in stages}Double logarithmic distribution of node degrees in stages.}
\end{center}		
\end{figure}

As shown in Table~\ref{Comparison of network central potentials with different linear relations}, first, it can be seen that relationships between stocks in the SSE A-shares market are still dominated by linear relationships, and the nonlinear relationship is relatively weak; the Pearson correlation is the most important method to describe the linear correlation, as it can better measure the intuitive and potential correlation relationship between stocks. Second, the linear causal and multi-stage dynamic threshold Pearson correlation networks of multiple The change trend of centrality potential is basically the same, the most important purpose of our study is to focus on the characteristics of HUB nodes, the change in trends of centrality potential of different methods to construct the network is consistent, which illustrates the robustness of the method. Finally, in the multi-stage dynamic threshold correlation network, the selection of threshold value also considers the filtering of redundant information through the connected subset, the change trend of centrality potential of each stage is more significant, which can better highlight the differences in centrality at different stages and discover the most closely correlated nodes from the financial contagion perspective. It can be seen that the reliability of the network construction method in this paper can be seen from the perspectives of the core linear relationship measure, the consistency of the central potential trend change, and the prominence of the financial status of HUB nodes. The CDF distribution of the network degree and the double logarithmic coordinate plot \cite{clauset2009power} are used to verify whether each phase conforms to the power-law distribution. The network is found to have a more significant thick-tail feature in the bear phase, as shown in Figures~\ref{CDF of node degrees in stages} and \ref{Double logarithmic distribution of node degrees in stages}.

As shown in Figures~\ref{CDF of node degrees in stages} and \ref{Double logarithmic distribution of node degrees in stages}, we obtain the corresponding probability distribution by the taking logarithm of the degree of the dynamic correlation network node of A-shares and note the following:
\begin{enumerate}

\item For most periods of the SSE A-share dynamic correlation network, the image as a whole is skewed to the right. Most nodes have a relatively small degree, and the number of HUB nodes with the largest degree is small, which is more in line with the power-law distribution and close to the scale-free network.

\item There are leading HUB stocks in the bull market stage. The probability of nodes with high degrees is relatively low, which indicates that there are a few HUB nodes with leading effects, and regulators can control those HUB stocks to suppress the phenomenon of overheating in the stock market.

\item During the bear market stage, the difficulty in risk control increases. The bear market node degree has a fat-tailed distribution. In BEAR 3, even through the degree distribution presents a U-shape, the probability of the fat-tailed distribution is abnormally high, indicating that there are more nodes closely related to each other. Together they caused the overall market to decline, and therefore increase the difficulty of risk management.
	
\end{enumerate}


\subsection{Analysis of meso-Industry of dynamic correlation network of SSE A-Shares }

\subsubsection{Dynamic Evolution of Small-World Effects in Industrial Boards}

In this section, the stock nodes are divided into the 11 sectors identified by the Global Idustry Classification Standard (GICS) using the full stock sample network at each stage. As, the telecommunication sector only contains two nodes, the focus is on analyzing the changes in the topological nature of the networks for the other ten industrial sectors. The topological nature of the meso-industry is based on a sub-industry study of the full sample network. We examine the stock's synergy within each industry sector and the differences in the sector's contribution to systemic risk, taking into account the differences in the number of nodes in each sector and the relative centrality.

\begin{table}

\centering
\caption{Comparison of small-world effects within industry.}
\label{Comparison of small-world effects within industry}
\small

\begin{tabular}{lcccccc}
	\toprule
	GICS sector & \multicolumn{2}{c}{BULL 1}                                     & \multicolumn{2}{c}{BEAR 1}                                     & \multicolumn{2}{c}{BULL 2}                                     \\ \cmidrule(l){2-7} 
	& \multicolumn{1}{c}{Cluster}     & \multicolumn{1}{c}{Average} & \multicolumn{1}{c}{Cluster}     & \multicolumn{1}{c}{Average} & \multicolumn{1}{c}{Cluster}     & \multicolumn{1}{c}{Average} \\
	& \multicolumn{1}{c}{coefficient} & \multicolumn{1}{c}{length}  & \multicolumn{1}{c}{coefficient} & \multicolumn{1}{c}{length}  & \multicolumn{1}{c}{coefficient} & \multicolumn{1}{c}{length}  \\ \midrule
	Real estate                  & 0                               & 1.333                       & 0.774                           & 1.656                       & 0.728                           & 1.663                       \\
	Consumer discretionary      & 0.351                           & 3.000                           & 0.762                           & 1.635                       & 0.772                           & 1.679                       \\
	Industrials                     & 0.478                           & 2.853                       & 0.734                           & 1.729                       & 0.67                            & 1.822                       \\
	Utilities                    & 0.507                           & 2.043                       & 0.676                           & 1.678                       & 0.734                           & 1.782                       \\
	Financial                      & 0.778                           & 1.333                       & 0.722                           & 1.839                       & 0.798                           & 1.673                       \\
	Energy                       & 0.583                           & 1.625                       & 0.755                           & 1.750                       & 0.855                           & 1.358                       \\
	Consumer staples             & 0.375                           & 1.667                       & 1.792                           & 1.792                       & 0.792                           & 1.681                       \\
	Information   technology     & 0.498                           & 2.287                       & 0.850                           & 1.389                       & 0.763                           & 1.672                       \\
	Health care                  & 0                               & 1.727                       & 0.722                           & 1.764                       & 0.689                           & 1.818                       \\
	Materials                & 0.504                           & 2.649                       & 0.733                           & 1.682                       & 0.720                           & 1.685                       \\ \midrule
	GICS sector                 & \multicolumn{2}{c}{BEAR 2}                                     & \multicolumn{2}{c}{BULL 3}                                     & \multicolumn{2}{c}{BEAR 3}                                     \\ \cmidrule(l){2-7} 
	& \multicolumn{1}{c}{Cluster}     & \multicolumn{1}{c}{Average} & \multicolumn{1}{c}{Cluster}     & \multicolumn{1}{c}{Average} & \multicolumn{1}{c}{Cluster}     & \multicolumn{1}{c}{Average} \\
	& \multicolumn{1}{c}{coefficient} & \multicolumn{1}{c}{length}  & \multicolumn{1}{c}{coefficient} & \multicolumn{1}{c}{length}  & \multicolumn{1}{c}{coefficient} & \multicolumn{1}{c}{length}  \\ \midrule
	Real estate                  & 0.808                           & 1.637                       & 0.522                           & 2.036                       & 0.796                           & 1.629                       \\
	Consumer discretionary      & 0                               & 1.286                       & 0                               & 2.817                       & 0.840                           & 1.721                       \\
	Industrials                     & 0.642                           & 2.199                       & 0.317                           & 3.945                       & 0.858                           & 1.456                       \\
	Utilities                    & 1.000                          & 1.000                       & 0.493                           & 2.57                        & 0.825                           & 1.349                       \\
	Financials                      & 0.800                           & 1.500                       & 1.000                            & 1.000                        & 0.750                           & 1.778                       \\
	Energy                       & 1                               & 1.000                        & 0.821                           & 1.556                       & 0.749                           & 1.747                       \\
	Consumer staples             & 0                               & 1.944                       & 0.542                           & 3.132                       & 0.850                           & 1.550                       \\
	Information   technology     & 0.460                           & 1.889                       & 0.750                           & 1.533                       & 0.862                           & 1.483                       \\
	Health care                  & 0                               & 1.500                       & 0                               & 1.000                        & 0.816                           & 1.547                       \\
	Materials                & 0.625                           & 2.822                       & 0.362                           & 3.273                       & 0.807                           & 1.565                       \\ \bottomrule
\end{tabular}	

\end{table}

The aggregation coefficient and the average shortest path are shown in Table~\ref{Comparison of small-world effects within industry}.
First, as a whole, the results are consistent with the characteristics of a macro network structure. The bear market in the abnormal fluctuation stage is stronger than the small-world connectivity effect in the general fluctuation stage. The asymmetric effect of the bull and bear periods is also significant within the industry. Table~\ref{Comparison of small-world effects within industry} shows that the small-world effect of each industrial sector in the bull stage is quite different; the clustering coefficient of each industry in the bear stage is above $ 0.67 $, the average path length is less than $ 1.84 $, the small-world effect of each industry becomes stronger, while the difference between industries is slight.

Second, the differentiation of internal characteristics within boards of directors is more obvious. Some sectors have maintained strong connectivity, while others have more internal connectivity fluctuations. The small-world effect is strong in the finance, energy, and utilities sectors and is relatively stable; the clustering coefficients fluctuate little with the average path length, especially in the financial sector. In the two bull periods and the fluctuated bear period, there is still a strong small-world effect; this indicates that the small-world effect of boards in these three sectors is relatively stable during market conversion.

Finally, the small-world effects of industrials, consumer discretionary, healthcare, and materials sectors fluctuate with the market. These sectors show long average path lengths and low aggregation coefficients in the two bull and bear periods, especially in the consumer discretionary, industrials and health care sectors. The clustering coefficient is close to $ 0 $ in BEAR 2 and BULL 3, and the number of associated edges is minimal in the whole plate, while they are small and medium worldwide in all boards in the bear market period. The results show that the correlations between stock nodes in these four sectors are greatly affected by the market situation. When the market price rises rapidly, the price of a stock in the sector does not necessarily drive the price of other nodes in the same sector.


\subsubsection{Analysis of Internal centrality within Industrial Boards}

Further analysis of the evolution of the mean and heterogeneity of centrality within sectors allows us to explore the differences in the contribution of different sectors to network risk spillover and overall connectivity, as shown in Table~\ref{Comparison of the internal centrality within industry}.

\begin{table}

\centering
\caption{Comparison of the internal relative degree centrality within industry.}
\label{Comparison of the internal centrality within industry}

\begin{tabular}{lrrrrrr}
	\toprule
	GICS sector             & \multicolumn{2}{c}{BULL 1}                                          & \multicolumn{2}{c}{BEAR 1}                                          & \multicolumn{2}{c}{BULL 2}                                          \\ \cmidrule(l){2-7} 
	& \multicolumn{1}{c}{Mean} & \multicolumn{1}{c}{Hetero.} & \multicolumn{1}{c}{Mean} & \multicolumn{1}{c}{Hetero.} & \multicolumn{1}{c}{Mean} & \multicolumn{1}{c}{Hetero.} \\ \midrule
	Real estate              & 0.244                          & 37.500\%                           & 34.024                         & 3.480\%                            & 41.098                         & 3.390\%                            \\
	Consumer discretionary  & 2.295                          & 4.210\%                            & 38.153                         & 1.660\%                            & 40.706                         & 1.600\%                            \\
	Industrials                 & 1.351                          & 3.130\%                            & 28.791                         & 1.000\%                            & 26.098                         & 1.010\%                            \\
	Utilities                & 4.435                          & 9.190\%                            & 35.484                         & 35.480\%                           & 32.056                         & 4.490\%                            \\
	Financial                  & 7.273                          & 28.130\%                           & 25.455                         & 10.460\%                           & 40.000                             & 12.190\%                           \\
	Energy                   & 5.882                          & 16.410\%                           & 10.294                         & 14.540\%                           & 31.618                         & 9.140\%                            \\
	Consumer staples         & 0.581                          & 14.840\%                           & 23.730                         & 3.130\%                            & 35.196                         & 2.760\%                            \\
	Information technology & 3.535                          & 7.840\%                            & 56.768                         & 2.680\%                            & 33.434                         & 3.180\%                            \\
	Health care              & 0.425                          & 18.000\%                           & 35.459                         & 2.860\%                            & 30.357                         & 3.030\%                            \\
	Materials            & 1.695                          & 3.590\%                            & 30.580                         & 1.250\%                            & 32.517                         & 1.200\%                            \\ \bottomrule
	GICS sector         & \multicolumn{2}{c}{BEAR 2}                                          & \multicolumn{2}{c}{BULL 3}                                          & \multicolumn{2}{c}{BEAR 3}                                          \\ \cmidrule(l){2-7} 
	& \multicolumn{1}{c}{Mean} & \multicolumn{1}{c}{Hetero.} & \multicolumn{1}{c}{Mean} & \multicolumn{1}{c}{Hetero.} & \multicolumn{1}{c}{Mean} & \multicolumn{1}{c}{Hetero.} \\ \midrule
	Real estate              & 9.512                          & 6.880\%                            & 1.220                          & 16.000\%                           & 31.829                         & 3.790\%                            \\
	Consumer discretionary  & 0.143                          & 14.000\%                           & 0.430                           & 6.440\%.                           & 37.694                         & 1.760\%                            \\
	Industrials                 & 0.532                          & 4.060\%                            & 0.715                          & 2.630\%                            & 47.856                         & 0.870\%                            \\
	Utilities                & 0.806                          & 21.880\%                           & 4.032                          & 9.130\%                            & 51.210                          & 3.890\%                            \\
	Financial                  & 29.091                         & 13.280\%                           & 12.727                         & 19.390\%                           & 25.455                         & 14.030\%                           \\
	Energy                   & 11.029                         & 16.670\%                           & 19.118                         & 11.830\%                           & 30.147                         & 8.920\%                            \\
	Consumer staples         & 0.581                          & 10.160\%                           & 2.104                          & 6.240\%                            & 45.283                         & 2.560\%                            \\
	Information technology & 0.909                          & 20.990\%                           & 0.808                          & 20.310\%                           & 44.343                         & 3.030\%                            \\
	Health care              & 0.255                          & 33.330\%                           & 0.340                          & 12.500\%                           & 45.918                         & 2.640\%                            \\
	Materials            & 1.325                          & 3.970\%                            & 0.655                          & 4.040\%                            & 39.695                         & 1.170\% \\ \bottomrule                           
\end{tabular}
	
\end{table}

As shown in Table~\ref{Comparison of the internal centrality within industry}, the mean value of the centrality in the network is significantly higher than that of the other stages during the period of the external subprime shock and internal deleveraging adjustment. The risk release period of abnormal fluctuation is highly correlated with risk synergy in A-shares. This synergy further enhances risk transmission and amplifies risk impact and loss effects. The sector also has an asymmetric impact on the bear and bull periods, the same fall effect of the industrial sector is stronger than the synchronic rise effect, and the same synchronic impact of each sector in the internal deleveraging stage in 2015 is stronger than that under the impact of the external subprime mortgage crisis.

Given the heterogeneity of the centrality in Table~\ref{Comparison of the internal centrality within industry}, the heterogeneity of the bull period is generally larger than that of the bear period. When the stock market rises, the status of stocks in the bull period is unequal. A few core nodes may lead their industry sector. By supervising a small number of stocks, the overheating tendency can be better managed. There is little difference in the status of stocks in the sector during a bear period and a high correlation between stocks. There is a risk that changes in any stock may infect the whole sector and therefore increase the difficulty of supervision.

By comparing the mean value of relative degree centralities and heterogeneity of stocks in various sectors, we show:

\begin{enumerate}

\item The heterogeneity of the financial sector and the energy sector centrality has little variation in different periods, and the mean value of degree centrality stays at a high level.

The heterogeneity of the financial industry is between $ 10\% $ and $ 20\% $ in most periods. On average, the heterogeneity of the financial sector is on the high side, which is related to the increasing influence of the main core nodes on the financial sector. The energy sector shares a similar pattern with the financial. The heterogeneity of the energy sector is higher in different periods, the difference between the bull and bear period is small, and the repetition rate of core nodes in each stage is high.

\item The heterogeneity of consumer discretionary, industrials and materials sectors in different stages of centrality heterogeneity changes very little, with the average being relatively low.

Apart from the BEAR 2 period, the heterogeneity of the whole network is low, the heterogeneity of the industrial plate is lower than $ 5\% $ in different periods, and there is no conspicuous control position in the sector. At the same time, the small-world effect of the industrial sector in the bear period is powerful, and changes to any node can easily lead to systemic risk. The heterogeneity of the materials sector is less than $ 5 $ in different stages. It rarely belongs to a HUB in any stage because the materials sector is the fundamental industry of the national economy--this presents difficulties for the regulatory authorities to carry out related regulation.

\item The heterogeneity of utilities, real estate, information technology, consumer staples, and health care sector present significant differences in different stages.
	
\end{enumerate}

The utility sector is different from other sectors. Its heterogeneity in BEAR 1 is $ 31\% $, far higher than the bull period, which is relatively high in all industries.

At the BULL 1 stage, the heterogeneity of the real estate sector is as high as $ 37.5\% $, while the heterogeneity of the bear period is about $ 3\% $, which is strongly affected by the economic cycle. In the long-term, the importance of each stock in the sector becomes smaller and is related to the excessive prosperity of the real estate sector in recent years, in which the core nodes leading the sector continue to vary.

The heterogeneity of the information technology sector is high in BEAR 2 and BULL 3, reaching $ 20\% $. There are only $ 45 $ network nodes in the information technology sector, but the core nodes change significantly. This relates to the fierce competition in the industry and the constant change of influence of the stock nodes.

Although the heterogeneity of the consumer staples sector is different under different market conditions, the high coincidence rate of core nodes in each stage is similar to real estate and information technology sectors. The market performance differences are also relatively large. The heterogeneity in BEAR 2 is about $ 33\% $.


\subsection{Analysis of Micro Node Status of SSE A-Shares Dynamic Correlation Network}

The scale-free nature of the macro structure of the network and the meso industry sector variability analysis shows that the dynamic correlation network of SSE A-shares is particularly prominent in the bull period in the leading nodes. The market HUB nodes play key roles in cross-sector and intra-sector connectivity changes; when the probability of HUB nodes in the bull market is low, the presence of a high number of nodes is beneficial for regulators to monitor for market overheating.\footnote{Due to the limited space, the results for ranking of centrality at different stages in 2005-2016 is available upon request.}

Examining the sector distribution of HUB nodes, the nodes with a higher degree of centrality in the fluctuating bear period are concentrated to three sectors: materials, real estate, and industrials sectors. The higher degree nodes in the internal analysis of materials and industrial sectors often have a higher degree of centrality in the whole A-shares stock connectivity network; for example, Xinjiang Joinworld, Redstar Developing, Zhongyuan Expressway, and Standard shares.

The industrial and materials sectors have strong control over other sectors and are in the communication bridge with market risk transmission. Strengthening supervision in these sectors can effectively slow down the spread of market risk. The analysis of high HUB nodes in the bull period and the non-standard characteristics of the network, the node with the highest connectivity status of the bull period in the A-shares market is significantly different from that of other stocks. As such, regulators can prevent market overheating through effective supervision of key industries and enterprises. As the degree distribution in bear periods has a thick tail, only a small number of high HUB nodes are concerned; this has little effect on early warning or preventing a of sharp market decline.

We then undertake a regression analysis of influencing factors of $ 50\% $ and $ 75\% $ nodes in each stage, respectively, calculated following Equation~\eqref{equation}.

\begin{equation}
\label{equation}
\begin{split}
	\text{Centrality}_i =& \alpha + \beta_{1} \text{Current ratio}_i + \beta_{2} \text{Quick ratio}_i + \beta_{3} \text{Leverage}_i
	\\& + \beta_{4} \text{Turnover rate}_i + \beta_{5} \text{ROE}_i + \beta_{6} \text{Market value}_i.
\end{split}	
\end{equation}

The dependent variable is the degree centrality of each node, and the degree centrality of stock $ i $ is estimated based on the network constructed at each stage. The independent variables are the current ratio, quick ratio, gearing ratio, total asset turnover ratio, return on net assets and total stock market value of listed company $ i $ at that stage. The data at each stage are cross-sectional data. The coefficients are estimated by the multivariate quadratic assignment method, using $ 1000 $ random permutations; this can overcome the variable covariance problem. The financial situation of listed companies is often considered to be one of the key concerns of investors, which often directly affects the trading and activity of stocks and thus their correlations; The total market capitalization of listed companies is the most direct reflection of the size of stocks; large-cap stocks are often directly related to the trend of the composite index, and thus tend to be at the center of the market \cite{agrawal2019does,li2020patent}. This leads to the following hypotheses:

\begin{itemize}
\item[Hypothesis 1:] The better the overall financial status of listed companies, the more attention they receive from investors, and therefore the higher their centrality status.

\item[Hypothesis 2:] The higher the market capitalization of a listed company, the greater its impact on the SSE A-shares index and the more it receives investors' attention, and therefore its centrality status is higher.	
\end{itemize}

We therefore add the financing and financing quota of each stock in Equation~\eqref{equation}. The financing and financing business of SSE A-shares started on 31 March 2010, therefore there was no financing and financing business in the first three periods of our study, and was only piloted in the fourth period of our study. So for the latter two stages, the model of whether to introduce financing and financing quota is designed respectively. Financing and financing business is usually considered to enhance market liquidity; therefore, stocks with higher financing and financing quota may be more liquid and receive more investors' attention, calculated for the last two periods of our study using Equation~\eqref{equation2}.

\begin{equation}
\label{equation2}
\begin{split}
	\text{Centrality}_i =& \alpha + \beta_{1} \text{ROE}_i +\beta_{2} \text{Market value}_i + \beta_{3} \text{Current ratio}_i +\beta_{4} \text{Quick ratio}_i 
	\\& + \beta_{5} \text{Leverage}_i +\beta_{6} \text{Turnover ratio}_i + \beta_{7} \text{Security Financing}_i.
\end{split}	
\end{equation}

This leads to the followning assumptions:
\begin{itemize}

\item[Hypothesis 3:] Listed companies have higher financing and financing limits, their trading and liquidity are higher, and therefore their centrality is higher.

\end{itemize}

Since we mainly focus on the characteristics of listed companies with high-degree ranks, while some securities with low-degree ranks tend to create noise effects due to inactive trading, we take the sample of stocks with degree centrality ranking in the top $ 50\% $ and $ 75\% $ by truncating the tails. The final phased parameter estimation results are shown in Tables~\ref{Results of influencing factors of node degree top 0.5} and \ref{Results of influencing factors of node degree top 0.75}.


\begin{table}

\centering
\caption{Results of influencing factors of node degree with top $ 50\% $ degree nodes.}
\label{Results of influencing factors of node degree top 0.5}
\footnotesize

\begin{tabular}{lccccccc}
	\toprule
	Stage & LR & QR & LEV       & TATO     & ROE         & Value & Yrading \\ \midrule
	BULL 1           & -0.1452         & 0.3102      & -0.0103 & -0.2559 & -0.0038   & -0.0079      & -                         \\
	BEAR 1           & -1.6452         & 2.6453**    & -0.0747 & -0.0747 & 1.4660    & 0.0764**     & -                         \\
	BULL 2           & 0.2275          & -1.2055     & -0.0421 & -0.5449 & -0.0071   & -0.0272      & -                         \\
	BEAR 2           & 0.1807***       & -0.2040     & 0.0046* & -0.1919 & 0.0000*** & 0.0047***    & -                         \\
	BULL 3 (1)       & 0.0124          & -0.0825     & -0.0060 & -0.1563 & -0.0084   & 0.0011       & -                         \\
	BEAR 3 (1)       & 0.3226          & -0.0714     & 0.0154  & -0.5984 & -0.0001   & 0.0241**     & -                         \\
	BULL 3 (2)       & 0.0097          & -0.0786     & -0.0059 & -0.1525 & -0.0083   & 0.0014       & -0.0057                   \\
	BEAR 3 (2)       & 0.2818          & -0.0018     & 0.0200  & -0.4811 & 0.0015    & 0.0223*      & 0.0279 \\ \bottomrule                   
\end{tabular}		

\end{table}

\begin{table}

\centering
\caption{Results of influencing factors of node degree with top $ 75\% $ degree nodes.}
\label{Results of influencing factors of node degree top 0.75}
\footnotesize

\begin{tabular}{lccccccc}
	\toprule
	Stage & LR & QR & LEV       & TATO     & ROE         & Value & Yrading \\ \midrule
	BULL 1     & -0.0567               & 0.1786               & -0.0091**             & -0.2238**              & -0.0009                & -0.0066**                      & -                           \\
	BEAR 1     & 1.1014               & -1.8978               & -0.1634                & 2.7026                 & -0.0856**             & 0.0106                         & -                           \\
	BULL 2     & 2.5937**             & -4.2816               & -0.0344                & -3.1095                 & 0.0296                & -0.0782**                      & -                           \\
	BEAR 2     & 0.1038                & -0.1132***           & 0.0046                & -0.1503**              & 0.0001                & 0.0068                         & -                           \\
	BULL 3 (1) & 0.0348               & -0.0905               & -0.0019                & -0.1338                 & -0.0093***            & 0.0074                         & -                           \\
	BEAR 3 (1) & -0.2373               & 1.1915               & 0.0167                & 0.5285                 & -0.0136                & 0.0123                         & -                           \\
	BULL 3 (2) & 0.0371               & -0.1096**            & -0.0026                & -0.0921**              & -0.0003                & 0.0249                         & -0.0034                    \\
	BEAR 3 (2) & 1.2886               & -0.4256               & -0.0744                 & 0.0095                 & -0.0607***            & 0.0015*                        & 0.2854 \\ \bottomrule                   
\end{tabular}			

\end{table}

To test Hypothesis 1, we use the company's financial situation, mainly from the liquidity, leverage and return on equity and other key indicators of the company's financial status to analyze the impact of the company's operations on its stock market relevance centrality. Overall, there are several financial indicators that affect the company's centrality in the network. Bear period variables are more significant than bull period variables as, overall, stocks are more actively traded in bull periods. While financial variables are more significant in bull periods than bear periods. The decline in BEAR 1 is due to the impact of the external subprime crisis, and the decline in BEAR 3 is due to domestic regulatory deleveraging. The common feature of these two phases is driven by BEAR 2, which is macro in nature and the result of a shock decline following a bull period; the liquidity ratio, leverage and return on equity of listed companies have different degrees of influence on their centrality in this phase.

The market capitalization size of listed companies, which is the focus of Hypothesis 2, is the most significant variable in each phase. Although there are a few leading nodes in the bull period, the influence and characteristic factors of the top $ 50\% $ of nodes are not significant. The leading nodes in the bear period show thick-tail, while the factors that influence the top $ 50\% $ of nodes have severval significant factors--therefore, different regulatory strategies should be formulated for bull periods and bear periods. HUB nodes prevent market overheating, and bear periods focus on the general characteristics of the high number node, where the market capitalization variable is significantly positively correlated with the degree in all bear periods, i.e., large-cap companies are at the center of the market correlation in the decline, while in the bull period it shows a weak and negative correlation, which should be a key concern.

When studying HUB nodes, we found they stand out in bull periods and have a low probability distribution, with only a few points being high degree numbers. Analysis of the market capitalization of top $ 10 $ stocks with the highest degrees shows they tends to be small and mid-cap stocks as they tend to have higher liquidity in bull markets and establish correlations with more stocks. When expanding the sample to nodes ranked in the top $ 50\% $ of degree numbers, degree numbers still show a weak. In constrast, in the bear periods, the market capitalization indicator has maintained a significant positive correlation in BEAR 1, BEAR 2, BEAR 3 and the model after adding the financing index. The distribution of the high degree nodes in the bear periods shows a thick tail; therefore, the top $ 10 $ degree nodes cannot be simply taken as the top 10 nodes in the bear period. Overall, the leading stocks in bear periods are mainly large-cap companies; this is very different from the bull periods, so regulators should ensure they distinguish between bull and bear periods when developing policies.

Hypothesis 3 focuses on the impact of financing and financing on market volatility during the BULL 3 and BEAR 3 stages. It is generally believed that highly leveraged lending plays a key role in the sharp rise and fall of these two phases. The on-course financing and financing business is supervised by the Securities Regulatory Commission; the amount of on-course financing and financing is announced daily in a timely manner, providing a high degree of information transparency. The off-course matching business is conducted mainly through various shadow banks, which enter the market through unregulated channels and are not supervised. 
In the BULL 3 and BEAR 3 stages, a stability test is conducted by introducing the amount of financing and financing securities, and replacing the total amount of financing and financing securities with the difference of financing and financing securities, the mean value, the amount of financing, the amount of financing securities and the ratio of total amount of financing and financing securities to market capitalization in this stage. In the process of high leverage aggregation and rapid deleveraging, the opaque nature of over-the-counter (OTC) financing rapidly accumulates bubbles during short-term blind rises. This creates financial risks, affecting market sentiment during rapid falls, causing panic, and subsequently leading to a sharp rise and fall in the whole market. The whole market appears to surge and plunge.

Overall, through the regression estimates of the samples in the top $ 50\% $ and $ 75\% $ of the degree, market capitalization is the most significant variable, and the significance remains consistent and robust in different stages of the two samples. The sign of the variables that are significant in financial indicators also remains consistent in the two samples. However, the financing and financing funds that are more strictly regulated in the market are not significantly related to the central status of the stock; rather, the different types of OTC matching funds driven by OTC shadow banks have a greater impact on the abnormal fluctuations of the stock market due to the lack of supervision.


\subsection{Comparision with Granger Linear and Nonlinear Networks}

We compare the dynamic correlation networks of the SSE A-shares market with two types of causality networks, generated by Granger linear and nonlinear methods, to verify the results reported above. This section gives a summary of the main findings, while detailed results are provided in \ref{Appendix A}.

The standard Granger causality test \citep{granger1980testing, billio2012econometric} is adopted to describe the linear before-after relationship between two stocks, where $ p_{ij} $ is the p-value of the hypothesis test that stock $ j $ does not cause stock $ i $. We define the adjacency matrix as:
\begin{equation}
	a_{ij} = \left\{ \begin{array}{cc}
		1,   &   p_{ij}  < \theta_1;   \\
		0,   &   p_{ij}  > \theta_1.   \\
	\end{array} \right.	
	\label{Eq Adjacency Matrix p-value}
\end{equation}
where $ \theta_1 $ is the given significant level. We also discuss the Granger nonlinear connections in the SSE A-shares market and follow previours scholars \cite{hiemstra1994testing, hmamouche2020nlints} by calculating p-values and then using Equation~\eqref{Eq Adjacency Matrix p-value} to build nonlinear Granger networks.

\begin{figure}
\begin{center}
	\includegraphics{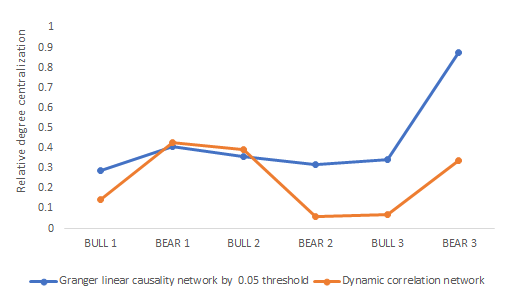}
	\caption{\label{Fig Comparison of relative degree centralizations} Comparison of relative degree centralizations in the dynamic correlation and Granger linear causality network. The blue (orange) line represents the Granger linear (dynamic correlation) network when the threshold is set to be $ 0.05 $ ($ 0.6456 $). }	
\end{center}						
\end{figure}

Table \ref{Tab Densities of Granger Nonlinear Nets} demonstrates that Granger nonlinear causality relationships are not the dominant interaction patterns because the densities of the Granger nonlinear causality networks are only around $ 0.1656\% $ under different thresholds over all periods except for BEAR 2 (about $ 0.34\% $). Such sparse network structures provide convincing evidence of mainly considering linear relationships in the SSE A-shares market over the entire period.

One of our primary concerns is the dynamic features of HUB nodes identified by the changes in degrees. As shown in Figure~\ref{Fig Comparison of relative degree centralizations}, the relative degree centralizations of the dynamic correlation and Granger linear causality network suggest that degrees in two different types of networks share a very similar trend over the entire period. Table~\ref{Tab Granger Linear Nets Relative Degree Centralization} in \ref{Appendix A} further support this argument by reporting centralizations under different thresholds in Granger linear causality networks, implying the consistence of degrees in correlation and causality networks. Moreover, compared with centralizations in Granger linear causality networks, these topological indicators in the dynamic correlation network show more significant changes over the entire period and highlight the structural differences in bull and bear periods in the markets, leading to the discovery of HUB stocks in the market.


\section{Conclusion}

This paper divides the SSE A-shares data from 2005 to 2016 into six different phases based market index, policy and volatility. We construct a dynamic correlation network of SSE A-shares by improving the threshold determination method through the maximum connected sub-graph and $3\sigma$ principle iteration. We compare it with the multi-stage A-shares dynamic network constructed by linear and non-linear causality methods. We then compare the differences in network structures between abnormal volatility and calm periods with sharp rises and falls in the market, and identify the difference between abnormal volatility and calm periods from different dimensions. The main findings are as follows.

First, the reliability of the multi-stage dynamic threshold method constructed in this paper is illustrated by comparing the multi-method construction network. In general, the market is more connected during an abnormal stock market volatility phase than a calm period, the correlation among stock nodes is closer, the heterogeneity among nodes becomes smaller, and the node degrees are relatively convergent. At the same time, in both abnormal volatility phases, the process of capital allocation and leverage-regulatory deleveraging in 2015 had a greater impact on the SSE A-shares market than the subprime mortgage crisis. This is mainly because the A-shares market was not closely linked to foreign capital flows in 2008, while the 2015 was a liquidity shock caused by rapid domestic deleveraging.

Second, from the perspective of industry differentiation, the small-world
effect in the financial, energy and utilities sectors is significant across all phases, with a high degree of linkage between stocks within sectors and high rates of intra-sector risk contagion; the industrial, consumer discretionary, healthcare, and materials sectors are more affected by the market and the small-world effect is unstable. In a bear period, investors should pay more attention to managing risk in investments in sectors with faster intra-sector risk contagion; they should also consider diversification in investments to avoid intra-sector and inter-sector risks due to the strong correlation among related sectors.

Third, from the stock node status, there are more prominent HUB nodes in bull periods; the average distance between most core nodes is not short; risk contagion to the whole network takes a period of time; if regulation is timely and effective, it can effectively prevent the market overheating. In bear periods, node degrees have thick-tails showing a stronger effect of the same fall, and large-capitalization stocks play a greater role in market linkage. Regulators should pay attention to the difference between bull and bear periods to formulate regulatory policies.

Overall, market microstructures in bull and bear periods are asymmetric, particularly in terms of volatility \cite{bentes2018is,bai2019economic,xiao2018asymmetric}. In this paper, based on the cross-stage variability of SSE A-shares market stock price trends, earnings volatility and macro policies, we construct a multi-stage dynamic correlation network of SSE A-shares using different methods and find that SSE A-shares have asymmetry in both bull and bear markets in terms of macro network connectivity and aggregation
characteristics, and inter-sector risk contagion, and node status characteristics. This not only provides evidence for investors to manage risk and regulators to formulate policy, but also provides more supporting evidence for the asymmetric characteristics of capital markets during bull and bear periods in market microstructure theory.


\newpage
\appendix

\section{ }
\label{Appendix A}

\begin{table}[h]

\centering
\caption{Densities ($ \times 10^{2} $) of SSE A-shares Granger nonlinear causality networks under different thresholds}
\label{Tab Densities of Granger Nonlinear Nets}

\begin{tabular}{crrrrrrr}
	\toprule
	Stage  & \multicolumn{7}{l}{Threshold value}                                                                                                                                                       \\ \cline{2-8} 
	& \multicolumn{1}{c}{0.2} & \multicolumn{1}{c}{0.15} & \multicolumn{1}{c}{0.1} & \multicolumn{1}{c}{0.075} & \multicolumn{1}{c}{0.05} & \multicolumn{1}{c}{0.025} & \multicolumn{1}{c}{0.01} \\	 \hline
	BULL 1 & 0.1658                  & 0.1656                   & 0.1656                  & 0.1656                    & 0.1656                   & 0.1656                    & 0.1656                   \\
	BEAR 1 & 0.1656                  & 0.1656                   & 0.1656                  & 0.1656                    & 0.1656                   & 0.1656                    & 0.1656                   \\
	BULL 2 & 0.1656                  & 0.1656                   & 0.1656                  & 0.1656                    & 0.1656                   & 0.1656                    & 0.1656                   \\
	BEAR 2 & 0.3590                  & 0.3519                   & 0.3462                  & 0.3426                    & 0.3391                   & 0.3352                    & 0.3322                   \\
	BULL 3 & 0.1656                  & 0.1656                   & 0.1656                  & 0.1656                    & 0.1656                   & 0.1656                    & 0.1656                   \\
	BEAR 3 & 0.1656                  & 0.1656                   & 0.1656                  & 0.1656                    & 0.1656                   & 0.1656                    & 0.1656                   \\ \bottomrule
\end{tabular}	
	
\end{table}


\begin{table}[h]

\centering
\caption{Relative degree centralizations of SSE A-shares Granger linear causality networks}
\label{Tab Granger Linear Nets Relative Degree Centralization}

\begin{tabular}{crrrrrrr}
	\toprule
	Stage  & \multicolumn{6}{l}{Threshold   value}                                                                                                                                                      & Dynamic corr.               \\ \cline{2-7} 
	& \multicolumn{1}{c}{0.2} & \multicolumn{1}{c}{0.15} & \multicolumn{1}{c}{0.1} & \multicolumn{1}{c}{0.075} & \multicolumn{1}{c}{0.05} & \multicolumn{1}{c}{0.025} & \multicolumn{1}{l}{network} \\ \hline
	BULL 1 & 0.2752                  & 0.2854                   & 0.2923                  & 0.2923                    & 0.2899                   & 0.2697                    & 0.1462                      \\
	BEAR 1 & 0.5170                  & 0.5121                   & 0.4846                  & 0.4599                    & 0.4089                   & 0.2954                    & 0.4258                      \\
	BULL 2 & 0.4566                  & 0.4582                   & 0.4048                  & 0.3766                    & 0.3586                   & 0.3031                    &  0.3921                      \\
	BEAR 2 & 0.4250                  & 0.4116                   & 0.3895                  & 0.3526                    & 0.3204                   & 0.2460                    & 0.0590                      \\
	BULL 3 & 0.3714                  & 0.3726                   & 0.3696                  & 0.3595                    & 0.3449                   & 0.3050                    & 0.0698                      \\
	BEAR 3 & 0.6943	                 & 0.7465            &	0.8046                        & 0.8371                    & 0.8738                   & 0.9178	                 & 0.3370  \\ \bottomrule
\end{tabular}	
	
\end{table}

\clearpage
\bibliographystyle{elsarticle-num-names} 
\bibliography{ReferenceDynamicCorrelationNets}






\end{spacing}		
	
\end{document}